\documentstyle[11pt]{article}
\title{Magnetic and spin evolution of neutron stars
       in close binaries}
\author{V.Urpin$^{1,2)}$, U.Geppert$^{3)}$ and D.Konenkov$^{1)}$ \\
     1) A.F.Ioffe Institute of Physics and Technology,\\
        194021 St.Petersburg, Russia \\
     2) Department of Mathematics,
        University of Newcastle,\\
        Newcastle upon Tyne NE1 7RU, UK \\
     3) Astrophysikalisches Institut Potsdam,\\
        An der Sternwarte 16,
        D-14482 Potsdam, Germany}

\date{Jule 10, 1997}

\begin{document}

\maketitle

\begin{abstract}
The evolution of neutron stars in close binary systems with a 
low-mass companion is considered assuming the magnetic field
to be confined within the solid crust. We adopt the standard
scenario of the evolution in a close binary system in accordance with 
which the neutron star passes throughout four evolutionary phases
("isolated pulsar" -- "propeller" -- accretion from the wind of a 
companion -- accretion due to Roche-lobe overflow). Calculations
have been performed for a great variety of parameters characterizing
the properties both of the neutron star and low-mass companion. We
find that neutron stars with more or less standard magnetic field 
and spin period being processed in low-mass binaries can evolve to 
low-field rapidly rotating pulsars. Even if the main-sequence life 
of a companion is as long as $10^{10}$ yr, the neutron star can
maintain a relatively strong magnetic field to the end of the
accretion phase. The considered model can well account for the
origin of millisecond pulsars. 


\end{abstract}

\section{Introduction}

In recent years evidences have been obtained for a relatively
fast magnetic field decay in neutron stars undergoing accretion
in binary systems (see, e.g., Taam \& van den Heuvel 1986,
Bhattacharya \& van den Heuvel 1991). The weak surface magnetic
fields of many binary radiopulsars ($B \leq 10^{9}$G), for which
an efficient accretion has taken place, strongly suggest the 
idea of the accretion induced magnetic field decay in these neutron
stars. Most of the low-magnetic-field pulsars are believed to be
old neutron stars experienced a mass transfer in low-mass binaries
where accretion ensues due to Roche-lobe overflow and may last
relatively long.

Recently Geppert \& Urpin (1994) and Urpin \& Geppert (1995) have
suggested a simple and efficient mechanism of a rapid field decay 
in neutron stars undergoing accretion. Accretion changes drastically
the thermal evolution of the neutron star if the accretion rate is
sufficiently high and a mass transfer lasts a sufficiently long
time (see, e.g., Fujimoto et al. 1984). The energy release due to 
accretion heats the neutron star and reduces the crustal conductivity
since the latter depends on the temperature. A decrease of the 
conductivity accelerates the ohmic decay of the magnetic field if 
this field is maintained by currents in the crust. If the mass
transfer phase lasts as long as $10^{6}-10^{7}$ yr the surface field
strength can be reduced by a factor $\sim 10^{2}-10^{4}$ depending
on the magnetic configuration at the beginning of the accretion phase.

Accretion influences not only the field decay but the spin evolution
of pulsars as well, and the strength of the magnetic field is one
of the dominating factors that determines this evolution. If the
total amount of the accreted matter is large enough ($\geq 0.01 - 0.1
M_{\odot}$), the neutron star in a low-mass binary can be "recycled"
to millisecond periods (Alpar et al. 1982). 

The accretion phase, however, does not exhaust all possible stages
in the evolution of neutron stars entering low-mass binary systems. 
The evolution of such systems is extremely long since even the 
main-sequence lifetime of a low-mass companion exceeds 
$10^{9}$ yr. In accordance with the standard scenario, the neutron
star in a close binary can pass throughout several evolutionary
phases (see, e.g., Pringle \& Rees 1972, Illarionov \& Sunyaev 1975):

1) the initial obscured radio pulsar phase in which the pressure of 
the pulsar radiation is sufficient to keep plasma of the companion 
stellar wind away from the neutron star magnetosphere; the radio
emission of the pulsar is partly (or completely) absorbed by the
plasma cloud surrounding the binary system thus the pulsar is
practically unobservable; the pulsars spins down due to the 
magnetodipole radiation;

2) the propeller phase in which the radiation pressure reduced due 
to the field decay and spin down cannot prevent the infalling matter
from an interaction with the magnetosphere but the spin rotation is
sufficiently rapid yet to eject the wind matter due to the 
centrifugal force and to transfer the angular momentum from the
neutron star to the wind matter;

3) the wind accretion phase in which the matter from the wind falls
on to the surface of the neutron star; nuclear burning of the
accreted material heats the neutron star and causes the accretion-driven
field decay; in the course of this phase the neutron star can slightly
spin up since the accreted matter carries a certain amount of the
angular momentum;

4) the enhanced accretion phase which starts when the secondary star
leaves the main-sequence and fills its Roche-lobe; a heavy mass
transfer on to the neutron star heats the interior to a very high
temperature $\sim 10^{8}-10^{8.5}$K (Fujimoto et al. 1984, 
Miralda-Escude et al. 1990) and accelerates the accretion-driven
field decay; a steady Keplerian disc is formed by the accretion flow
outside the neutron star magnetosphere and the accretion torque spins
up the neutron star to a very short period.    

The neutron star processed in the above evolutionary picture can
probably prolong its life as a radiopulsar with relatively low magnetic
field and short spin period after accretion due to Roche-lobe overflow 
is exhausted. The properties of pulsars produced due to evolution in 
low-mass binaries have not been analysed in detail yet and we address
this problem in the present paper. Note that a similar problem has
been considered by Jahan Miri \& Bhattacharia (1994) who assumed that
the magnetic evolution is determined by the expulsion of the field
lines from the neutron star core due to interaction between the
fluxoids in the proton superconductor and the vortices in the neutron
superfluid. The authors, however, restricted themselves to the case 
of wide low-mass binaries and considered the evolution only during
the main-sequence life time.

The paper is organized as follows. We describe in detail the adopted
model in Section 2. The results of calculations are presented in
Section 3. In Section 4, we discuss our results and compare them with 
the available observational data on the evolution of neutron stars
in low-mass binaries. 

\section{Description of the model}

We consider the evolution of the neutron star in a binary system with 
the secondary being a low-mass star. The binary is assumed to be 
relatively close thus the companion can fill the Roche lobe after
the end of its main-sequence life. Both the magnetic and spin evolution
of the neutron star may be strongly influenced by accretion from the
companion. During its main-sequence evolution (which lasts as long as
$(1 - 5) \times 10^{9}$ yr), the secondary loses mass due to the 
stellar wind and a certain fraction of the wind plasma can generally
be captured and accreted by the neutron star. After the secondary
leaves the main-sequence and fills the Roche-lobe, the rate of mass 
loss and, hence, the accretion rate can be drastically enhanced.

We assume that the magnetic field of the newly born neutron star is
created by an unspecified mechanism either in the course of collapse
or soon after the neutron star is born. We also assume that the field 
is maintained by currents in the crust and occupies originally only
a fraction of the crust volume. The evolution of the crustal
field is determined by the conductive properties of the crust and
material motion throughout it (see for a detail Sang \& Chanmugam 1987, 
Urpin \& Geppert 1995). Neglecting the anisotropy of conductivity,
the induction equation governing the evolution of the magnetic field
reads
$$
\frac{\partial \vec{B}}{\partial t} = - \frac{c^{2}}{4 \pi}
\nabla \times \left( \frac{1}{\sigma} \nabla \times \vec{B}
\right) + \nabla \times ( \vec{v} \times \vec{B}) \;, 
\eqno(1)
$$
where $\sigma$ is the conductivity and $\vec{v}$ is the velocity
of material movement. The velocity $\vec{v}$ is caused by the flux 
of the accreted matter throughout the crust and may be non-zero only
during those evolutionary phases when accretion on to the neutron
star surface is allowed. The total amount of the accreted mass in 
most cases does not exceed 0.1-0.2$M_{\odot}$ (see Bhattacharya \&
van den Heuvel 1991), even for neutron stars undergoing a very
extended accretion phase. Therefore, the mass and radius of the 
neutron star do not change significantly and, hence, the corresponding
changes in the crustal thickness, density profile etc. are also
negligible. The total mass flux is the same throughout the crust
and should be taken equal to the accretion rate. Assuming spherical
symmetry of the mass flow in deep crustal layers, one obtains the
radial velocity of material movement
$$
v = \frac{\dot{M}}{4 \pi r^{2} \rho} \;, \eqno(2)
$$
where $\dot{M}$ is the accretion rate and $\rho$ is the density. 
This inward directed flow tries to push the magnetic field into the 
deep crustal layers and, generally, can influence the evolution of
the field at a high accretion rate.

Introducing the vector potential for a dipole magnetic field
$\vec{A} = (0,0, A_{\varphi})$, where $A_{\varphi} = s(r,t) \sin 
\theta /r$ and $r, \theta, \varphi$ are the spherical coordinates,
the induction equation (1) can be transformed to 
$$
\frac{\partial s}{\partial t} = \frac{c^{2}}{4 \pi \sigma}
\left( \frac{\partial^{2} s}{\partial r^{2}} - \frac{2s}{r^{2}}
\right) - v \frac{\partial s}{\partial r} \;. \eqno(3)
$$
At the surface $r=R$, the standard boundary condition for a dipole
field should be fulfilled, i.e. $R \partial s/\partial r + s = 0$.
We assume the core of the neutron star to be superconductive, thus
the magnetic field cannot penetrate into the core. Therefore, the
second boundary condition is $s=0$ at the crust-core boundary which
lies at the density $2 \times 10^{14}$ g/cm$^{3}$.

The conductive properties of the solid crust are determined by 
scattering of electrons on phonons and impurities (see, e.g., Yakovlev
\& Urpin 1980). Scattering on phonons dominates the transport 
processes at a relatively high temperature and low densities, whereas 
scattering on impurities gives the main contribution to the conductivity 
at a low temperature and large densities. The total conductivity of 
the crystallized region is given by
$$
\frac{1}{\sigma} = \frac{1}{\sigma_{ph}} + \frac{1}{\sigma_{imp}} \;,
\eqno(4)
$$
where $\sigma_{ph}$ and $\sigma_{imp}$ are the conductivities due
to electron-phonon scattering and electron-impurity scattering
mechanisms, respectively. The phonon conductivity depends on the
temperature, decreasing when $T$ increases, and hence heating caused 
by accretion accelerates the field decay. The impurity conductivity
is practically independent of the temperature but its magnitude is 
determined by the so called impurity parameter,
$$
Q = \frac{1}{n_{i}} \sum_{n'} n'(Z-Z')^{2} \;, \eqno(5)
$$
where the dominant background ion species has the number density
$n_{i}$ and charge $Z$, and $n'$ is the number density of an impurity
species of charge $Z'$; the summation is over all species of impurities. 
In our calculations, we use the numerical data for the phonon
condutivity obtained by Itoh et al. (1993) and a simple analytical
expression for the impurity conductivity derived by Yakovlev \& Urpin
(1980). Since the conductivity depends on the temperature, the
magnetic evolution of the neutron star is strongly sensitive to its 
thermal history. According to the scenario suggested in Section I,
we divide the evolution into four essentially different phases: \\
the "isolated pulsar" phase (phase I) in which the neutron star
practically does not feel the influence of its companion, \\
the "propeller" phase (phase II) in which, due to a fast spin rotation, 
the neutron star ejects the wind matter penetrating into a magnetosphere, \\
the wind accretion phase (phase III) in which accretion of the wind
plasma is allowed, and \\
the accretion phase (phase IV) in which the secondary fills its
Roche-lobe and the neutron star experiences a heavy mass transfer. \\  

Evidently, that during the phases I and II the thermal evolution
of the neutron star in a binary system exactly follows that of an
isolated neutron star. For the time dependence of the temperature
during these two phases we use the standard cooling model as given 
by Van Riper (1991) for 1.4$M_{\odot}$ neutron star with 
the Pandharipande-Smith equation of state. During the phases
III and IV, however, the neutron star can be substantially heated
by accretion. The thermal evolution of accreting neutron star has 
been a subject of study of several papers (see, e.g., Fujimoto et
al. 1984, Miralda-Escude et al. 1991, Zdunik et al. 1992). It was
argued that, after a relatively short initial stage, the 
temperature distribution within the neutron star reaches a steady
state. Then, the temperature is practically uniform in deep layers
of the crust of neutron star with standard neutrino emissivities.
The temperature that the crust will attain in the steady state
depends on the accretion rate, $\dot{M}$. For the phase III with a low
accretion rate, we use the crustal temperature computed by Zdunik
et al. (1992) for the accretion rate within the range $10^{-11}
\geq \dot{M} \geq 10^{-16} M_{\odot}$/yr. For the phase IV when
accretion is caused by Roche-lobe overflow, the internal temperature
has been calculated by Fujimoto et al. (1984) for the accretion rate 
$\dot{M} = 3 \times 10^{-10}M_{\odot}$/yr and by Miralda-Escude et
al. (1991) for $\dot{M}$ within the range $10^{-10} \geq \dot{M}
\geq 10^{-11}M_{\odot}$/yr. We matched their results and slightly
extrapolate the obtained dependence in order to estimate $T$ for
$\dot{M} \sim 10^{-9}M_{\odot}$/yr, the maximal accretion rate we
use in calculations.

Note that the accretion rate during the phases III and IV is
sensitive to the separation between the stars and their masses.
In the present study we will not consider the dependence of the
evolution on the orbital period of a binary despite of the
importance of this dependence for interpretation of observational
data (see van den Heuvel \& Bitzaraki 1995). We are planning to
address this problem in a forthcoming paper.

The spin evolution of the neutron star being strongly coupled 
with its magnetic evolution differs essentially for
different evolutionary phases. As it was mentioned, the neutron
star is practically not influenced by a companion during the phase
I since the wind plasma of the secondary is stopped by the pressure
of magnetodipole radiation behind the light cylinder. The stopping
radius, $R_{s}$, is determined by the balance between the
dynamical pressure of the wind and the radiative pressure of
magnetodipole waves. The dynamical pressure of the wind is of the
order of $\rho_{w} V_{w}^{2}$ where $\rho_{w}$ and $V_{w}$ are
the density and velocity of the wind plasma, respectively, near
the stopping radius. The density, $\rho_{w}$, at the distance
$s$ from the secondary may be estimated as $\rho_{w} \approx
\dot{M}_{0}/ 4 \pi s^{2} V_{w}$ with $\dot{M}_{0}$ being the rate
of mass loss of the secondary. The pressure of the magnetodipole
radiation at the distance $R_{s}$ from the neutron star is 
$\sim (1/4 \pi R_{s}^{2})(B^{2} R^{6} \Omega^{4}/6 c^{4})$ where
$B$ is the surface field strength at the magnetic pole and $\Omega$
is the spin angular velocity of the neutron star. Equating this pressure to the dynamical pressure of the wind at the nearest to the
neutron star point and assuming that the stopping radius is
smaller than separation between stars, $a$, i.e. $s \sim a$, 
one obtaines the expression for $R_{s}$,
$$
R_{s} \sim \left( \frac{B^{2} R^{6} \Omega^{4}}{6 c^{4} V_{w}}
\cdot \frac{a^{2}}{\dot{M}_{0}} \right)^{1/2} \;. \eqno(6)
$$
In our calculations, it is more convenient to use the accretion
rate, $\dot{M}$, instead of the rate of mass loss $\dot{M}_{0}$.
At $a > R_{G}$ where $R_{G} \approx 2GM/V_{w}^{2}$ is the radius 
of gravitational capture (Bondi 1952) and $M$ is the neutron star 
mass, these quantities are related by $4 \dot{M}/R_{G}^{2} \approx 
\dot{M}_{0}/a^{2}$; we assume that the wind velocity is larger 
than both the sound velocity of the wind plasma and the velocity of 
orbital motion. The phase I lasts while the stopping radius is larger 
than the radius of gravitational capture, $R_{s} > R_{G}$.

With the assumption that the spin down torque on the neutron
star is caused by radiation of the magnetodipole waves (Ostriker \&
Gunn 1969), the evolution of the spin period, $P$, is governed by
$$
P \dot{P} = \frac{8 \pi^{2} B^{2} R^{6}}{3 c^{3} I} \;, \eqno(7)
$$
where $I$ is the moment of inertia. Note that approximately the same 
braking law applies if the torque is determined by currents in
the model suggested by Goldreich and Julian (1969). At $R_{s} < R_{G}$, 
the pressure of the magnetodipole radiation cannot prevent the wind
plasma from a penetration behind the light cylinder. The influence 
of the wind plasma on the spin-down rate becomes appreciable
and the further spin evolution of the neutron star departs from that
of an isolated pulsar. In our simplified model, the condition
$R_{s} = R_{G}$ determines the transition between the phases I and II.

During the phase II, the wind matter can directly interact with the 
neutron star magnetosphere. However, rotation is rather fast yet and 
the magnetosphere acts as a "propeller",
ejecting the infalling wind matter. During this phase, a certain 
fraction of the spin momentum of the neutron star is transfered to 
the matter being ejected, and the spin period may increase a few 
orders of magnitude. In calculations presented here, we adopt a 
very simplified model of the interaction between the accretion flow 
and magnetosphere which is, nevertheless, qualitatively close
to that commonly encountered in the literature (see, e.g., Pringle
\& Rees 1972, Illarionov \& Sunyaev 1975). We assume that the accreted
matter interacts with the magnetosphere at the so called Alfven
radius, $R_{A}$,
$$
R_{A} = \left( \frac{R^{6} B^{2}}{4 \dot{M} \sqrt{GM}} \right)^{2/7}
\approx 1.2 \times 10^{9} \left( \frac{B_{12}^{2}}{\dot{M}_{-10}}
\right)^{2/7} \; {\rm{cm}}, \eqno(8)
$$
where $B_{12} = B /10^{12}$G and $\dot{M}_{-10} = \dot{M}/10^{-10}
M_{\odot}$yr$^{-1}$; the numbers are given for the
neutron star with $M=1.4M_{\odot}$ and $R=1.64 \times 10^{6}$ cm.
If the neutron star rotates rapidly and its angular velocity is
larger than the Keplerian angular velocity at the Alfven radius, 
$\Omega_{K}(R_{A})$, the wind matter has to be expelled by the
magnetosphere carrying out the spin angular momentum. In order
to be expelled, the wind matter should get the angular momentum 
a little bit larger than it corresponds to the Keplerian
rotation. Therefore, the amount of the spin momentum lost by
the neutron star per unit time, $\dot{J}_{p}$, may be estimated as
$$
\dot{J}_{p} \simeq 4 \pi R_{A}^{4} \rho_{w} \Omega_{K} v_{r}
\simeq \dot{M} \sqrt{GMR_{A}} \;, \eqno(9)
$$
where $v_{r}$ is the radial velocity of the accreting flow near
the magnetospheric boundary. The corresponding spin-down rate of 
the neutron star is 
$$
\dot{P} \simeq \frac{P^{2} \dot{J}}{2 \pi I} \simeq
\beta P^{2} B^{2/7} \dot{M}^{6/7} \;, \eqno(10)
$$
where $\beta = (GMR^{2}/8)^{3/7}/\pi I$. The neutron star rotation
slows down until the angular velocity becomes comparable to 
$\Omega_{K}(R_{A})$. We determine the critical period, $P_{eq}$,
at which the transition between the "propeller" and accretion
phases takes place from the condition $\Omega = \Omega_{K}(R_{A})$,
$$
P_{eq} = \frac{18.6 B_{12}^{6/7}}{\dot{M}_{-10}^{3/7}} \; s.
\eqno(11)
$$
This period can be rather large if the neutron star is accompanied
by a secondary star with a low rate of mass loss. Equation (11)
determines the so called spin up line in the $B$ - $P$ plane. If
the spin rotation slows down to the value given by equation (11),
it is generally accepted that the accreting matter can flow throughout
the magnetosphere and accrete on to the surface, thus the neutron 
star enters the phase III. During this phase accretion from the
stellar wind is allowed and, as it was mentioned, nuclear burning 
of the accreted material provides some heating of the neutron star
interior and induces the accretion-driven field decay. In its turn,
the decay of the magnetic field decreases the critical period, 
$P_{eq}$. Since accretion from the wind in a binary cannot be 
exactly spherical, the accreting matter carries some amount of the 
angular momentum to the neutron star (see Geppert et al. 1995) and,
due to this, the neutron star can experience spin up to a shorter
period. Probably, a balance may be reached in spin up and the
rate of the field decay, thus the neutron star slides down the
corresponding spin-up line (see, e.g., Bhattacharya \& Srinivasan
1991). This evolution lasts either until the end of the main-sequence 
lifetime of the companion (when it fills the Roche-lobe and the neutron
star enters the phase IV) or until the magnetic field becomes weak 
to such extent that the accreting angular momentum is insufficient to 
maintain a balance in spin up and the rate of the field decay. 

The amount of the angular momentum carried by the accreted matter
can be estimated from the condition at the magnetospheric
boundary assuming that the angular momentum there is characterized
by its Keplerian value, $\rho \Omega_{K} R_{A}^{2}$, multiplied
by a some "efficiency" factor $\xi$. This factor is $\sim 1$ if
the accreted wind matter forms a Keplerian disc outside the
magnetosphere. If the disc is not formed, then the efficiency of
spin up is much lower and the parameter $\xi$ is probably of
the order of 0.1 - 0.01 (see Jahan Miri \& Bhattacharya 1994). The
amount of the spin angular momentum added to the neutron star per
unit time, $\dot{J}_{w}$, can be estimated as
$$
\dot{J}_{w} \simeq \xi \cdot 4 \pi R_{A}^{4} \rho \Omega_{K} v_{r}
\simeq \xi \dot{M} \sqrt{GMR_{A}} \;. \eqno(12)
$$
The corresponding spin-up rate is
$$
\dot{P} \simeq - \xi \beta P^{2} B^{2/7} \dot{M}^{6/7} \;. \eqno(13)
$$
If the magnetic field becomes very weak, $\dot{P}$ may be small
and accretion cannot maintain a balance in the spin up and field decay 
rates, thus the neutron star leaves the spin-up line.

After the companion fills its Roche-lobe and the phase IV starts,
accretion onto the neutron star is strongly enhanced and heating
of the crust leads to a rapid field decay. In low-mass binaries,
accretion due to Roche-lobe overflow can probably last as long as
$(1-5) \times 10^{7}$ yr. During this phase, the accreted matter 
likely forms the Keplerian disc outside the neutron star magnetosphere. 
Therefore, the rate of the angular momentum transfer and the
spin-up rate are given by equations (12) and (13), respectively,
with the "efficiency" parameter $\xi \sim 1$. The neutron star
spins up in accordance with these equations until it approaches 
the spin-up line corresponding the enhanced accretion rate.
Probably, the accretion rate has to reduce when the neutron star
reaches the spin-up line (Bhattacharya \& Srinivasan 1991) but,
nevertheless, the field continues to decay due to the accretion-driven
mechanism. During the further evolution, a balance should be reached 
in spin up and the rate of the field decay like the case of accretion 
from the wind. Due to this balance, the neutron star slides down the
corresponding spin-up line with the rate which is determined by the
field decay. In reality, the evolution on the spin-up line may be
very complicated but, in our calculations, we will adopt a maximally
simplified model. Namely, we will neglect a possible difference in 
the accretion rate before and after the neutron star reaches the 
spin-up line. As mentioned above, the amount of matter accreted onto
the surface may be reduced when the star evolves on the spin-up line.
However, this decrease of the accretion rate cannot probably change
drastically the internal temperature because, at high accretion rates,
the temperature is close to saturation and depends relatively weakly 
on the accretion rate (see Miralda-Escude et al. 1990 and Fujimoto 
et al. 1984). Therefore, hopefully, our assumption does not change
substantially the rate of the magnetic field decay on a spin-up line.
Note that in our model, the neutron star slides down the corresponding
spin-up line with the maximal rate but, in reality, the evolution
will be slower. The star can leave the spin-up line either if accretion
is exhausted or if the field becomes too weak to maintain a balance
in spin up and the field decay.

\section{Numerical results and discussion}

We consider the evolution of the neutron star for a wide range of
accretion rates and initial magnetic fields. Calculations
are performed for the neutron star model with $M=1.4 M_{\odot}$ based 
on the Pandharipande-Smith equation of state (Pandharipande \& Smith
1975). This model is a representative of stiff equations of state,
therefore the radius and thickness of the crust for the corresponding
neutron star are relatively large, $\approx$16.4 km and $\approx$4.2 km, 
respectively. We choose for calculations one of the stiff models 
because the magnetic evolution of isolated neutron stars based on 
such models is in a better agreement with the available observational 
data on pulsar magnetic fields (Urpin \& Konenkov 1997).
The evolution of neutron star models with equations of state softer 
than that of Friedman-Pandharipande seem to be inconsistent with 
B-P distribution of radio pulsars.

The conductive properties of the crust depend on its chemical
composition which, unfortunately, is rather uncertain for neutron
stars. This dependence is not significant during the initial evolution 
when the magnetic field is confined to the layers with not very high 
density and the difference between existing models of the chemical 
composition is not very large. However, the effect of a composition 
may be important during the phase IV when the neutron star is essentially 
heated and the conductivity is determined by phonon scattering which 
is strongly sensitive to the chemical composition. Therefore, in our 
calculations, we adopt the simplest model in accordance with which the 
crust is composed of nuclei processed in various nuclear transformations 
in accreting neutron stars (see Haensel \& Zdunik 1990) during the whole 
evolution of the neutron star.

In calculations, the impurity parameter, $Q$, is assumed to be constant 
throughout the crust during the whole evolution and to range from 0.001 
to 0.1. The initial spin rotation is assumed to be relatively fast, 
$P_{0} = 0.01$ s. Note, however, that the results are not sensitive 
to the value of $P_{0}$. On the contrary, the evolution is strongly 
dependent on the original magnetic configuration of the neutron star. 
The initial magnetic field is assumed to be confined to the outer 
layers of the crust with densities $\rho \leq \rho_{0}$. The calculations 
presented here are performed for $\rho_{0}$ ranged from 
$4 \times 10^{11}$ to $10^{13}$ g/cm$^{3}$. These values correspond to 
the depth from the surface $\approx$820 and 1100 m, respectively. 
The choice of $\rho_{0}$ is imposed by comparing the calculated 
magnetic and spin evolution of isolated neutron stars with a crustal 
magnetic field with the available observational data on $P$ and 
$\dot{P}$ for radiopulsars (see Urpin \& Konenkov 1997). The initial 
field strength at the magnetic pole, $B_{0}$, is taken within 
the range $10^{13} \geq B_{0} \geq 10^{11}$ G.
        
The present study is mainly addressed to neutron stars entering binary 
systems with a low-mass companion. For such secondary stars, the 
main-sequence life-time, $t_{ms}$, may last as long as $10^{9}- 
10^{10}$ yr. During the main-sequence evolution of the secondary, the 
accretion rate onto the neutron star is determined by both the 
characteristics of stellar wind and separation between the stars, and 
may generally vary within a wide range. In calculations, we suppose 
$\dot{M} = 10^{-13}$, $10^{-15}$ and $10^{-17} M_{\odot}$/yr during the 
propeller and wind accretion phase; $V_{w} = 500$ km/s, $a= 5 \times 
10^{12}$ cm. We also assume $\xi = 0.1$ for the efficiency parameter. 
Note, however, that our calculations show a weak sensitivity of
the evolutionary tracks to the particular choice of $\xi$. After the 
secondary leaves 
the main-sequence and fills its Roche-lobe, the accretion rate has to 
be strongly enhanced, $\dot{M} \approx 10^{-9}-10^{-10} M_{\odot}$/yr. 
We do not consider a higher accretion rate because of the lack of 
calculations on the thermal structure of accreting neutron stars with 
$\dot{M} > 3 \times 10^{-10} M_{\odot}$/yr.   
  
Figures 1-4 show examples of the magnetic, spin and thermal 
evolution of the neutron star in a binary system for different 
values of the parameters. Numbers in the top panels indicate
the evolutionary phase. Note a difference in the horizontal scales 
for different phases. The evolutionary tracks are shown in 
solid lines if the accretion rates due to Roche-lobe overflow
is $\dot{M}=10^{-10}M_{\odot}$/yr and in dashed lines if $\dot{M}=
10^{-9}M_{\odot}$/yr.

%
%

In Fig.1, we plot the evolution for a relatively strong initial 
magnetic field, $B_{0} = 10^{13}$ G, and for $\rho_{0} = 10^{13}$ 
g/cm$^{3}$. The impurity parameter is $Q=0.01$, the duration of the 
main-sequence life-time of a low-mass companion is $3 \times 10^{9}$ yr.
Accretion due to Roche-lobe overflow is assumed to last as long as
$10^{8}$ yr thus the total amount of accreted mass can reach $0.1
M_{\odot}$ if $\dot{M}=10^{-9} M_{\odot}$/yr. During the phase III, 
the accretion rate is $10^{-15}M_{\odot}$/yr. This model gives an 
example of the neutron star being processed in all evolutionary stages. 
The phase I, when the evolution of the neutron star is not influenced 
by a companion, lasts a bit less than $10^{7}$ yr. During this short 
stage, the field does not decay significantly thus its surface strength 
is about $2 \times 10^{12}$ G when the neutron star enters the "propeller"
phase. An efficient spin down caused by a strong magnetodipole radiation 
leads to a rapid increase of the spin period from $10^{-2}$ s to 
$\approx 6$ s. During the next evolutionary stage when the neutron
star works as a propeller, the field continues to decay slowly 
because, after $10^{7}$ yr of the evolution, the temperature
falls down to a very low value and the conductivity of crustal
matter is high. Note that during this stage the conductivity is 
determined by the impurity scattering and depends on the impurity
parameter $Q$. The phase II is relatively long ($\approx 230$ Myr) 
but the field is still very strong ($\sim 10^{12}$ G) to the end of 
this phase. Due to such a strong field, the neutron star can eject
the wind matter penetrating into a magnetosphere as long as the spin
period is shorter than $\approx 4 \times 10^{3}$ s. When a spin 
rotation slows down to this period, the matter from the wind can fall 
on to the surface of the neutron star and it becomes a soft and not very
bright X-ray source. Nuclear burning of the accreted material heats 
the neutron star interior to $T \approx 2 \times 10^{6}$ K. However,
this heating is too weak to accelerate appreciably the field 
decay thus the magnetic evolution during the wind accretion phase
does not differ essentially from that of an isolated star. 
Note that the wind accretion can accelerate the decay appreciably 
only if the accretion rate is larger than $10^{-13} M_{\odot}$/yr.
To the end of the wind accretion, the field strength is still as
high as $2 \times 10^{11}$ G. The neutron star can slowly spin up 
getting some amount of angular momentum from the accreted wind 
matter but this amount is not enough to spin up to a short period, 
thus the period turns out to be as long as $\sim 10^{3}$ s after 
$3 \times 10^{9}$ when the companion ends its main-sequence life. 
The phase III is the most extended for the considered model, however, 
as it will be shown, it may not be the case for other models. 
The most dramatic changes are experienced by the neutron star during 
the phase IV when an enhanced accretion due to Roche lobe overflow 
starts. Accretion with the accretion rate $\dot{M}=10^{-10}-10^{-9} 
M_{\odot}$/yr or higher heats the neutron star to the temperature  
$T \approx (2-3) \times 10^{8}$ K and decreases substantially the
crustal conductivity. If the field is reduced only by a factor 
$\sim 50$ during the previous $3 \times 10^{9}$ years, the 
accretion-driven mechanism leads to a much more efficient decay. 
Thus, in the case $\dot{M} = 10^{-9} M_{\odot}$/yr, the field weakens 
$\approx 30$ times after $10^{8}$ yr of accreting when the 
neutron star accretes $0.1 M_{\odot}$. For a higher accretion rate, 
the decay is even stronger. Since the accreted matter carries out
a large amount of the angular momentum, the neutron star can accelerate 
its spin rotation to relatively short periods during the phase IV.  
The spin evolution is non-monotonous (this is shown in the
small panel in Fig.1). At the beginning of accretion, the neutron 
star spins up very rapidly and after $\sim 10^{4}$ yr it reaches 
the corresponding spin-up line. Since the magnetic field is still 
sufficiently strong ($\sim (1-2) \times 10^{11}$ G), the corresponding
critical period, $P_{eq}$, is relatively long ($\sim 1$ s). The  
further spin evolution proceeds slowly because the neutron star slides
down the spin-up line and the rate of this migration is determined by 
the field decay. Finally, when accretion is exhausted after $10^{8}$
yr, the neutron star processed in the all above transformations 
will work as a radiopulsar with a low magnetic field ($\sim (1-3)
\times 10^{10}$ G) and a short period ($\sim 0.04-0.3$ s). 

%

Fig.2 shows one more example of the evolution which is typical for
stars with a relatively weak initial magnetic field. The initial
field is assumed to be $10^{11}$ G, other parameters are the same 
as in Fig.1. In this case, the neutron star does not pass throughout 
all evolutionary stages: accretion of matter from the wind is impossible 
because rotation spins down very slowly due to a weak magnetic
field and, after $3 \times 10^{9}$ yr, the period is still shorter 
than the corresponding value of $P_{eq}$. A weak initial field results 
in a very long duration of the phase I for this model: the period 
slows down to a value $\sim 0.4$ s which allows the infalling matter 
to interact with the magnetosphere only after $\sim 600$ Myr. 
However, the magnetic field turns out to be reduced only by a factor 
$\sim 10$ and is $\sim 10^{10}$ G after this long phase. The rest 
of the main-sequence life-time of a companion (which is about 80\% of
this life-time), the neutron star works as a propeller ejecting the 
wind plasma. Despite of a long duration of this phase, both the 
magnetic field and spin rotation do not change significantly to the 
beginning of Roche-lobe overflow: the surface field strength decreases 
approximately by a factor 3, whereas the period increases from 0.4 to 
0.46 s. Note a great difference in the periods just before the accretion
phase for this model and the model plotted in Fig.1 where $P \approx
10^{3}$ s at $t = 3 \times 10^{9}$ yr. An enhanced accretion, however, 
can accelerate a spin rotation very rapidly thus an accreting neutron star 
forgets shortly about its period before the accretion phase. Recycling 
is particularly fast during first $\sim (4-10)$ Myr of accreting when 
the neutron star spins up to 10-30 ms depending on the accretion rate 
and approaches the corresponding spin-up line. The further spin 
evolution on the spin-up line proceeds slower but, nevertheless, the
star can reach millisecond periods if a mass transfer lasts as long as
$10^{8}$ yr. For this model, the "recycled" periods when accretion is 
exhausted are $\sim 1$ and $\sim 10$ ms for $\dot{M} = 10^{-9}$ and  
$10^{-10} M_{\odot}$/yr, respectively. The magnetic field is also 
substantially reduced during the phase IV under the action of the
accretion-driven mechanism. To the end of this phase, the surface 
field may be as low as $10^{8}$ G if $\dot{M}=10^{-9}M_{\odot}$/yr 
(when the total accreted mass is $0.1 M_{\odot}$) and $3 \times 10^{8}$ 
G if $\dot{M}=10^{-10}M_{\odot}$/yr (the accreted mass is $0.01
M_{\odot}$). When accretion exhausts, the neutron star can be observed 
as a millisecond radio pulsar with the period $\sim 1-10$ ms and 
magnetic field $\sim (1-3) \times 10^{8}$ G. 

%
%

In order to illustrate the effect of the impurity content, we plot 
in Fig.3 the evolution of the neutron star with a strongly polluted 
crust, $Q=0.1$. We also assume that for this model the field is
initially confined to the layer with a smaller density, $\rho_{0}=
10^{12}$ g/cm$^{3}$, corresponding to the depth from the surface 
$\approx 900$ m. Like the model plotted in Fig.1, this one passes 
also throughout all evolutionary stages. However, the field decays 
faster for this model because the crustal conductivity is lower due 
to a higher impurity content and the field is initially anchored in 
the layers with a lower density (and, hence, a lower conductivity).
A faster decay leads to slower spin down and longer duration of 
two first phases in comparison with Fig.1. Thus, the "isolated
pulsar" and "propeller" phases last $\approx 13$  and 650 Myr, 
respectively, whereas the duration of these phases in Fig.1 is
$\approx 6$ and 230 Myr. The maximal spin period which is reached 
to the end of the propeller phase does not exceed, however, $\approx
300$ s. An increase of the impurity parameter, $Q$, causes a particularly 
rapid field decay during the phases II and III, when the conductivity 
is determined by scattering on impurities. During this two phases, the 
field is reduced two orders of magnitude and is a bit lower than
$10^{10}$ G when the companion fills its Roche-lobe and an enhanced
accretion starts. Accretion of matter during $10^{8}$ yr leads to
a formation of a millisecond pulsar with the period within the
range $\approx 2-20$ ms and with the magnetic field $\approx (3-10)
\times 10^{8}$ G, depending on the accretion rate during the phase IV.

%
%

Fig.4 represents the evolution of the neutron star with an intermediate 
initial magnetic field, $B_{0}=10^{12}$ G, and with a longer main-sequence
lifetime of the companion, $t_{ms} = 5 \times 10^{9}$ yr. Other
parameters are the same as in Fig.1. The evolution does not differ
in the main from that plotted in Fig.1 except an essentially weaker 
magnetic field of the neutron star at the beginning of the accretion 
phase. For this model, $B \approx 2 \times 10^{10}$ G at $t=5 \times 
10^{9}$ yr. When accretion starts, such a weak magnetic field is 
probably unable to channel the inflowing matter towards the magnetic 
poles, thus the rotation of the neutron star will not cause the  
observed X-ray emission to have a pulsed appearence. Note that the 
same concerns also the models plotted in Fig.2 and Fig.3. On the 
contrary, the neutron star with the evolutionary scenario represented 
in Fig.1 can be observed (at least, at the beginning of the accretion 
phase) as a pulsating binary X-ray sources. The accretion-driven field 
decay causes a further decrease of the magnetic field thus, to the
end of the phase IV, it may be as weak as $10^{9}$ G if $\dot{M}=
10^{-9} M_{\odot}$/yr. If accretion lasts $10^{8}$ yr, the neutron 
star can be easily recycled to a very short period. For this model,
the final spin periods are $\approx 10$ and $100$ ms after accreting
$0.1$ and $0.01 M_{\odot}$, respectively. Note that the field
strength and period of neutron star processed in accordance with
the scenarios shown in Fig.2-4 are very similar to those of
millisecond pulsars.

Fig.5-10 show the evolutionary tracks of neutron stars in the B-P
plane for different values of parameters. We also plot in these
figures the observational data on isolated pulsars (dots) and pulsars 
in binary systems (open squares) taken from the catalogue by Taylor, 
Manchester \& Lyne (1993). The dotted lines show spin-up lines 
corresponding to different accretion rates (numbers near these lines 
indicate the logarithm of the accretion rate). The solid lines represent
the evolutionary tracks. All tracks are calculated for the case of 
accretion due to Roche-lobe overflow with $\dot{M}=10^{-9} M_{\odot}$/yr. 
Numbers near the tracks indicate the logarithm of the age, numbers
with the subscript (a) mark the time required for the neutron star 
to reach the corresponding spin-up line after the end of main-sequence 
evolution. Large crosses show the points on tracks where a transition 
takes place between the phases I and II, small crosses with the 
corresponding numbers mark the end of tracks after $10^{8}$ yr 
of an enhanced accretion.

%

Fig.5 plots the evolution of the neutron stars which accrete matter 
from the stellar wind with different accretion rates. Calculations
are performed for $\dot{M} = 10^{-12}$ (curve 1), $10^{-13}$ (2), 
$10^{-14}$ (3), $10^{-15}$ (4) and $10^{-17} M_{\odot}$/yr (5), the
duration of main-sequence lifetime of the companion is $3 \times 10^{9}$ 
yr. The initial field is $B_{0}=10^{13}$ G, the density penetrated 
by the initial field $\rho_{0} = 10^{12}$ g/cm$^{3}$, and 
the impurity parameter $Q=0.1$. The neutron stars starting the evolution 
with such parameters pass typically throughout all evolutionary 
stages. The only exception is the model 5 with a very weak stellar wind, 
$\dot{M} = 10^{-17} M_{\odot}$/yr. This wind is insufficient to slow 
down the spin period to the critical value (11) when accretion from
the wind becomes allowed and, therefore, the neutron star is still 
in the "propeller" phase when the companion fills its Roche-lobe and 
an enhanced accretion starts. The duration of phases is obviously 
strongly dependent on the rate of wind accretion. For instance, the
initial phase when the neutron star evolves like an isolated neutron 
star lasts as long as $\approx 160$ Myr if $\dot{M} = 10^{-17} 
M_{\odot}$/yr and only $10^{5}$ yr if $\dot{M} = 10^{-12} M_{\odot}$/yr. 
To the end of this phase, the magnetic field and period are ranged 
from $2 \times 10^{11}$ to $10^{12}$ G and from $6$ to $1$ s, 
respectively, for different models. After this phase, the field and 
spin are reduced sufficiently thus the wind matter can penetrate into 
the magnetosphere, and the neutron star works as a propeller. The 
"propeller" phase lasts until the neutron star moving to the right and 
down in the B-P plane reaches the corresponding spin-up line. Evidently, 
the star which experiences a stronger accretion from the wind requires 
a shorter time to approach the spin-up line. Thus, one has $P=P_{eq}$ 
after 5 and 630 Myr of evolution if $\dot{M}=10^{-12}$ and $10^{-15} 
M_{\odot}$/yr, respectively. As it was mentioned above, in the case of 
a very weak accretion with $\dot{M}=10^{-17} M_{\odot}$/yr the spin-up 
line can never be reached. Accretion from the wind starts when
the magnetic field is still rather high for the models 1-4, $B \sim
10^{11}-10^{12}$ G. This field is certainly able to channel the 
accreting matter towards the magnetic poles thus, probably, some of 
neutron stars accreting matter from the wind in low-mass binaries can 
be observed as weak pulsating X-ray sources. A further evolution of 
the neutron star on the spin-up lines is completely determined by the 
field decay which can be accelerated due to accretion induced heating 
for a relatively high accretion rate. The wind accretion phase is the 
most extended phase for all models plotted in Fig.5 with the exception 
of the model 5. The parameters of stars can be drastically changed
during this long phase if the accretion rate is sufficiently large.
For instance, the field is reduced by a factor $\sim 10^{4}$ and is
only $\sim 10^{8}$ G for the model 1 with the accretion rate $\dot{M}=
10^{-12} M_{\odot}$/yr. Note, however, that at the end of the 
main-sequence lifetime of the companion the magnetic fields of the 
models 2-5 which experience accretion with $\dot{M} \leq 10^{-13} 
M_{\odot}$/yr lie within a relatively narrow range $\sim (2-10) 
\times 10^{9}$ G despite of a large difference in a duration of 
different phases for these models. The main reason for this is a 
relatively weak heating caused by accretion with such low accretion rates 
what makes inefficient the accretion-driven mechanism at $\dot{M} 
\leq 10^{-13} M_{\odot}$/yr.

When the companion fills its Roche-lobe and an enhanced accretion 
starts, the neutron star spins up initially very rapidly: it requires 
only 0.2 Myr to reach the spin-up line for the model 5 and 1.6 Myr for 
the model 2. These time intervals are too short for an appreciable 
decay of the magnetic field, therefore, the corresponding pieces of 
tracks are represented by approximately horizontal plateaus. Note that 
the magnetic field of the model 1 is very weak after $3 \times 10^{9}$ yr 
of evolution ($\sim 10^{8}$ G), and this model will never reach the 
spin-up line. Like the case of accretion from the stellar wind, in its  
further evolution, the neutron star slides down the spin-up line and
the rate of this sliding is determined by the field decay. However,
in the case of Roche-lobe overflow, the accretion-driven mechanism
is much more efficient and, hence, the decay is much faster. During
the phase IV, the magnetic field decreases by a factor $\sim 30$ for
the models 2-5, thus, when accretion is exhausted, all these neutron 
stars will work as pulsars with weak magnetic fields $\sim (1-5)
\times 10^{8}$ G and short periods $\sim 2-10$ ms. The model 1
is also spun up to a millisecond period but its magnetic field
is substantially smaller than $10^{8}$ G.

%

In Fig.6, we plot the same as in Fig.5 but for the magnetic field 
confined initially more deep layers, $\rho_{0} = 10^{13}$ g/cm$^{3}$, 
and for the crust with a smaller impurity content, $Q=0.01$. Evidently, 
the magnetic evolution is much slower for a less polluted crust. Like the 
case shown in Fig.5, the models 1-4 pass throughout all evolutionary phases 
whereas the model 5 with a very weak accretion from the wind ($\dot{M}= 
10^{-17} M_{\odot}$/yr) avoids the phase III. Due to a slower field decay 
and, hence, a faster spin down, the duration of the phases I and II is 
shorter for the models in Fig.6 than for the models with the same wind in 
Fig.5. Therefore, the wind accretion starts when the magnetic field of the 
neutron star is very strong, $B \sim 2 \times 10^{12}$ G for all models. 
As in the previous case, the wind accretion is the most extended phase 
for the models 1-4. To the end of the main-sequence evolution of the 
companion ($t_{ms}=3 \times 10^{9}$ yr), the magnetic field of neutron 
stars which experience the wind accretion with a low accretion rate, 
$\dot{M} \leq 10^{-13} M_{\odot}$/yr, lies in a narrow range from $4 
\times 10^{10}$ to $3 \times 10^{11}$ G. Only accretion with $\dot{M} > 
10^{-13} M_{\odot}$/yr can substantially change the magnetic evolution 
during the phase III. 

Due to a higher magnetic field at $t= 3 \times 10^{9}$ yr, the models
plotted in Fig.6 reach the spin-up line on a much shorter time scale 
when an enhanced accretion starts. Thus, it requires only $4 \times
10^{3}$ yr for the model 5 and $\approx 1$ Myr for the model 1. Note
that, in the previous case, the model 1 could never reach the spin-up
line during the accretion phase. If accretion due to Roche-lobe overflow
lasts $10^{8}$ yr, the models 2-5 which experienced a weak accretion 
from the wind evolve to relatively weakly magnetized pulsars with
$B \sim (2-10) \times 10^{9}$ G rotating with the period $\sim 0.03-
0.1$ s. The model 1 with a very high accretion rate during the phase
III evolves to the millisecond pulsar with $P \approx 3$ ms and
$B \approx 10^{8}$ G. Obviously, the magnetic field and period are 
larger for all pulsars if the accretion phase is shorter. 

%
%

Fig.7 compares the evolution of neutron stars with different initial 
magnetic fields and with the same $\rho_{0}$ and $Q$. Calculations 
have been performed for $\rho_{0} = 10^{13}$ g/cm$^{3}$, $Q=0.01$ and 
$t_{ms}=3 \times 10^{9}$ yr. It is apparent from this figure that the 
evolution is strongly sensitive to the initial field strength. The
weaker is the magnetic field, the longer is the duration of the 
"isolated pulsar" and "propeller" phases and, correspondingly, the shorter 
is the wind accretion phase. For instance, the neutron star can work as 
an "isolated pulsar" only 0.5 Myr if $B_{0}=10^{13}$ G but this stage  
may be as long as 63 Myr if the initial field is weak, $B_{0}=10^{11}$ 
G. The strongly magnetized star with $B_{0}=10^{13}$ G can accrete 
matter from the stellar wind after $\approx 12$ Myr whereas the weakly 
magnetized one with $B_{0}=10^{11}$ G after $\approx 600$ Myr. To the 
beginning of the wind accretion, magnetic fields of the considered 
models lie within a wide range from $10^{10}$ to $2 \times 10^{12}$ G. 
The corresponding periods are $\sim 10-1000$ s. The wind accretion 
phase is the most extended phase for all models plotted in Fig.7. 
During two previous phases the magnetic field is typically reduced 
by a factor $\sim 5-10$, whereas the decrease during a wind accretion is about 
10-50 times. When the companion fills its Roche-lobe and a heavy mass
transfer starts, the field of the neutron star is $\approx 10^{9}$ and 
$\approx 4 \times 10^{10}$ G for the most weakly and most strongly 
magnetized among the models presented in Fig.7, respectively. Probably, 
these fields are too weak to channel the accreting matter towards the 
magnetic poles, thus the neutron star is unable to work as a pulsating 
X-ray source during the phase IV. An enhanced accretion accelerates the 
field decay thus, when the mass transfer is exhausted, the field and 
period of the models 1-3 turn out to be quite suitable for millisecond 
pulsars. The model 4 which has initially the strong magnetic field 
$B_{0}=10^{13}$ G may also be transformed into a weakly magnetized and
rapidly rotating pulsar to the end of the accretion phase. However, 
the field of this pulsar ($\sim 2 \times 10^{9}$ G) is a bit stronger
and the period ($\sim 30$ ms) is a bit longer than it is typical for 
millisecond pulsars. Evidently that, if accretion lasts longer than 
$10^{8}$ yr, the model 4 can also evolve to a millisecond pulsar with 
the standard parameters.

%
%

In Fig.8, the dependence of evolutionary tracks is shown on the duration 
of main-sequence lifetime of the companion, $t_{ms}$. Since we address 
mainly the evolution of neutron stars entering binary systems with a 
low-mass companion, calculations have been done for a sufficiently long 
$t_{ms}$, $10^{10} \geq t_{ms} \geq 10^{9}$ yr. Despite of the fact that 
we plot in this figure tracks for the neutron star with a relatively 
polluted crust with $Q=0.03$ and for the initial magnetic field $B_{0}= 
3 \times 10^{12}$ G which is certainly weaker than measured fields of 
many isolated pulsars, it is seen that crustal currents can maintain 
a sufficiently strong magnetic field during the whole evolution. This 
conclusion is in contrast with the widely accepted opinion that neutron 
star models with the crustal magnetic field are not suitable to account 
for a relatively strong magnetic field of many long-lived pulsars. Even 
for a "polluted" crust, the decay turns out to be very slow thus the 
neutron star can possess the field stronger than $10^{9}$ G after 
$10^{10}$ yr of evolution and an enhanced accretion.

During the "isolated pulsar" phase which lasts $\approx 20$ Myr, 
the field weakens by a factor $\approx 5$ whereas during the "propeller" 
phase which is as long as $\approx 630$ Myr the field is reduced only 
by a factor 2. Working during $\approx 610$ Myr as a propeller, the
neutron star spins down to the period of the order of $10^{3}$ s 
which is slow enough to accrete the wind matter. At this 
moment, the field is rather strong, $B \approx 3 \times 10^{11}$ G, 
thus the accreted matter has to be channeled towards the magnetic poles. 
Perhaps, during the wind accretion phase such a neutron star can be
observed as a weak and pulsating X-ray source. When the companion 
fills its Roche-lobe, the field of the neutron star can range from
$3 \times 10^{10}$ to $2 \times 10^{11}$ G depending on the duration
of main-sequence lifetime. The corresponding spin periods lie within 
the range 100-500 s. Due to a relatively strong magnetic field, a 
spin-up is initially very fast thus the neutron star approaches the 
spin-up line shortly after the beginning of Roche-lobe overflow. The 
field does not change practically during this short time interval
($\approx 10^{4}-10^{5}$ yr). In the course of the further evolution 
alongside the spin-up line, the field is decreased by a factor 
$\approx 30$ under the action of the accretion-driven mechanism 
thus to the end of the accretion phase a pulsar can be formed with 
$B \sim (2-5) \times 10^{9}$ G and $P \sim 20-70$ ms. If accretion 
lasts longer than $10^{8}$ yr and the total amount of accreted mass is 
larger than $0.1 M_{\odot}$, all the models plotted in Fig.8 can evolve 
to "standard" millisecond pulsars with $B \sim 10^{8}-10^{9}$ G and
$P \sim 1-10$ ms. Note that if the mass of a low-mass companion is 
relatively large thus the main-sequence lifetime is closer $10^{9}$ yr 
rather than $10^{10}$ yr, then the field at the beginning of the phase 
IV is sufficiently strong ($B \sim 2 \times 10^{11}$ G) to channel the 
accreted matter towards the magnetic poles, and the neutron star 
can work as a pulsating X-ray source. To the end of accretion, the
field has to be reduced to a very low value insufficient to produce 
X-ray pulses but sufficient for a neutron star to work as a radiopulsar. 

%
%

Fig.9 shows the dependence of the evolution on the impurity "pollution" 
of the crust. Calculations have been performed for the values of the 
impurity parameter, $Q$, within the range from 0.001 to 0.3. The 
duration of the main-sequence lifetime of the companion is $3 \times
10^{9}$ yr. During the initial evolution, the field decay does not 
practically depend on the impurity parameter. This is due to the fact 
that the neutron star is relatively hot initially and the crustal 
conductivity is determined by phonon scattering which is independent 
of $Q$. Later on, when the star cools down and the impurity scattering 
dominates the conductivity, the latter becomes sensitive to a "polution" 
of the crust and the decay may be essentially different for different $Q$.

The behaviour of all models plotted in Fig.9 is more or less standard
during the main-sequence evolution of the companion. 
The only exception is the model 1 with extremely large $Q$. Due to a
large impurity content and, as a consequence, a low conductivity, the
magnetic field of this model decays too fast. During the phases I and
II which last together $\approx 1.6 \times 10^{9}$ yr, the field strength has
to be reduced more than 1000 times and, when the neutron star approaches
the spin-up line corresponding to the wind accretion rate $\dot{M} = 
10^{-15} M_{\odot}$/yr, its field is too low ($B \sim 2 \times 10^{9}$
G) to maintain a balance in spin up and the rate of the field decay.
Therefore, the neutron star does not slide down the spin-up line in its 
further evolution but moves down the B-P plane much faster. To the 
beginning of Roche lobe overflow the field of the model I weakens to a 
very low value $\sim 10^{8}$ G. An enhanced accretion causes a further 
field decay thus when accretion is exhausted the field strength of this
model is extremely low ($\sim 10^{6}$ G) and the pulsar is practically unobservable.

The models 2-5 with $Q= 0.001-0.1$ indicate more or less reasonable  
behaviour. Eventually, the neutron star passes throughout all 
evolutionary phases and, after all transformations, it forms a pulsar 
with a weak magnetic field ($B \sim (2-30) \times 10^{8}$ G) and a short 
period ($P \sim 5-40$ ms). Note that, at the end of accretion, the
field of the model 5 turns out to be lower than that of the models 3
and 4 despite of a smaller impurity parameter for the model 5. Due 
to a substantially higher conductivity, the field of the model 5 
diffuses inward very slowly thus, after $3 \times 10^{9}$ yr, the
current maintaining the magnetic configuration is located at much lower
densities than for the models 3 and 4. When accretion due to Roche-lobe 
overflow heats the neutron star to the temperature $\sim 2 \times 10^{8}$
K, the conductivity of layers occupied by the current becomes essentially 
lower for the model 5 because of a low density. Therefore, the field
decays much faster for this model during the accretion phase and, at
the end of evolution, it is even lower than for the model 3 and 4. 

%
%

The dependence of the evolutionary tracks on the initial depth 
penetrated by the field is shown in Fig.10. We plot the tracks for 
four values of $\rho_{0}$: $10^{11}$ (curve 1), $10^{12}$ (2), $10^{13}$
(3) and $10^{14}$ g/cm$^{3}$ (4). Evidently, the smaller is the
initial depth penetrated by the field, the faster is the decay.
For the model 1 with the magnetic field anchored initially in a 
relatively narrow surface layer with the depth $< 600$ m, the decay 
is so rapid that the neutron star cannot even be processed in all 
evolutionary transformations: after $3 \times 10^{9}$ yr of evolution 
the rotation is still fast enough to prevent accretion from the
stellar wind. Due to this, the model 1 evolves from the "propeller"
phase directly to the enhanced accretion phase missing the wind
accretion. However, even for this model, the field at the end of the
mass transfer is sufficiently strong thus when accretion is exhausted
the neutron star can work as a millisecond pulsar with $B \sim 3 \times
10^{8}$ G and $P \sim 8$ ms. The models 2-4 pass throughout all phases
and, evidently, the final magnetic fields for them are higher and the
periods are longer. For example, at the end of accretion, $B$ may be 
as high as $\sim 10^{11}$ G for the model 4 where the field occupies 
initially about half a crustal thickness.

\section{Conclusion}

We examined the magnetic and spin evolution of neutron stars in close 
binary systems with a low-mass companion. Due to a long lifetime of 
the companion, evolutionary transformations of the neutron star may 
last as long as $10^{9}-10^{10}$ yr and may be very complex. Likely, 
most of neutron stars in such binary systems experiences a heavy mass 
transfer ensued due to Roche-lobe overflow. This mass transfer can be 
extremely extended and the total amount of the mass accreted by the 
neutron star can exceed $0.1-0.5 M_{\odot}$ (see, e.g., van den Heuvel
\& Bitzaraki 1995). Besides, the evolution of neutron stars in low-mass 
binaries has to be influenced by the stellar wind during the 
main-sequence life of the companion. Due to this, the neutron star in 
the course of its evolution passes throughout several essentially 
different evolutionary phases. 

Calculations presented here are performed in agreement with the standard 
scenario of evolution of the neutron star in a close binary system 
(Pringle \& Rees 1972, Illarionov \& Sunyaev 1975). According to 
this scenario, the neutron star can be processed in four main 
evolutionary phases. Initially, when the star rotates rapidly and the 
field is strong, the pressure of the pulsar radiation does not allow 
the wind matter to penetrate into the magnetosphere, and the neutron 
star evolves like an isolated pulsar. The duration of this stage may 
vary depending on the initial field strength and the rate of field 
decay as well as on the intensity of the stellar wind of the companion.
The initial phase is obviously shorter for a neutron star with a higher 
magnetic field and for a stronger wind. Thus, for the initial field as 
strong as $10^{13}$ G, the phase I may last only $\sim 0.1-1$ Myr if 
the neutron star captures gravitationally $\sim 10^{-12}-10^{-13} 
M_{\odot}$/yr from the wind plasma. However, this phase may be 
extremely long ($\sim 100-300$ Myr) if the initial field is weak, $B_{0} 
\sim 10^{11}$ G or if the rate of gravitational capture is small, $\sim
10^{-15}-10^{-17} M_{\odot}$/yr. For the most of considered models, 
the field is typically reduced by a factor $\sim 5-10$ during this 
phase except the cases when its duration is extremely long.

When the energy loss due to magnetodipole radiation slows down 
rotation to such extent that the wind matter can interact with the 
magnetosphere, the neutron star begins to work as a propeller ejecting
the wind plasma. Obviously, the propeller action is more efficient for
a stronger magnetic field. The "propeller" phase is 
usually more extended than the
phase I. During the "propeller" phase, the magnetic evolution of
the neutron star in a binary does not differ from that of an isolated
star. Since the star is relatively cool to the beginning of the phase 
II, the conductivity of the crustal matter is determined by impurity 
scattering and may be high for a low impurity content. Due to this,
typically, the field decays less during a more extended "propeller" phase 
than during a shorter "isolated pulsar" stage. The spin evolution 
proceeds, however, more rapidly because the rate of angular momentum 
loss is larger for the propeller effect than for the magnetodipole 
braking. During the "propeller" phase the neutron star can spin down
to a very long period $\sim 10^{2}-10^{4}$ s. Note that this period
may be shorter for stars with low magnetic fields and for a stronger
stellar wind.

In the further evolution, the spin period of the neutron star can 
reach the critical value given by equation (11). When rotation becomes
so slow, the centrifugal force is unable to eject the wind plasma
penetrating into the magnetosphere, and the matter falls on to the 
neutron star surface. Nuclear burning of the accreted material heats 
the neutron star, decreases the crustal conductivity and accelerates 
the field decay. Note that an efficiency of the accretion-driven 
mechanism of the field decay is strongly sensitive to the accretion rate. 
The effect of accretion is relatively small for a low accretion rate, 
$\dot{M} < 10^{-13} M_{\odot}$/yr, but it may be of a great importance 
for higher accretion rates. During the wind accretion, the neutron star 
slides down the corresponding spin-up line and the rate of this sliding 
is determined by the rate of the field decay. Depending on the duration
of this phase and the rate of accretion, the field can be reduced by a 
few orders of magnitude. Since the accreted wind matter carries
a some amount of the angular momentum, the neutron star can also
spin up slightly during this phase but, for the most of considered 
cases, the spin period is still longer 1 s to the end of the 
main-sequence evolution of a companion. Usually, the wind accretion 
phase is the most extended among all evolutionary phases. Note, 
however, that the star can miss this stage in some cases (low rate of 
mass loss by a secondary star, rapid decay of the magnetic field), 
thus the evolution goes directly from the "propeller" phase to an 
enhanced accretion.

In low-mass binaries, the neutron star experiences the most dramatic
changes during a heavy mass transfer which starts when the companion
ends the main-sequence evolution and fills its Roche-lobe. Probably, a
mass transfer in such systems can last as long as $10^{7}-10^{8}$ yr.
Nuclear burning of the accreted material can heat the neutron star
interior to the temperature $T \sim (2-3) \times 10^{8}$ K and
substantially reduce the crustal conductivity. Due to this, the 
accretion-driven mechanism of the field decay is especially efficient 
during the phase IV. Typically, the field may be reduced about 20-30 
times if accretion lasts $10^{8}$ yr, but the decrease may be larger
for a longer accretion. Depending on the initial magnetic 
configuration and the duration of the main-sequence evolution,
the surface magnetic field can fall down to the value $\sim 10^{8}-
10^{9}$ G to the end of a mass transfer. Since the matter accreted 
from the disc carries a large angular momentum, the spin evolves 
very rapidly. The neutron star reach a spin-up line corresponding 
to the enhanced accretion on a short time scale, $\sim 0.01-1$ Myr.
In its further evolution, the star moves along the spin-up line while 
the magnetic field enables to maintain a balance in spin up and 
the rate of the field decay. During this phase, the accretion torque 
spins up the neutron star to a very short period, $\sim 1-100$ ms.

Our calculations show that the neutron star with the crustal magnetic
configuration can maintain a relatively strong field during the whole 
evolution in low-mass binaries. Even if the main-sequence life of a 
companion lasts as long as $10^{10}$ yr and, after that, the neutron star 
experiences a strong accretion with $\dot{M} \sim 10^{-9} M_{\odot}$/yr
during $10^{8}$ yr, the magnetic field can be sufficiently strong 
to the end of the accretion phase. When accretion is exhausted, such
a neutron star works as a radio pulsar with a weak magnetic field, 
$\sim 10^{8}-10^{10}$ G, and a short period, $P \sim 1-100$ ms. These
parameters are close to those of millisecond pulsars and, perhaps, these 
objects are formed in accordance with the considered scenario. Note
that if the main-sequence evolution is shorter and a mass transfer is
not so extended, the neutron star processed in the above evolutionary
transformations should have a stronger magnetic field and a longer
period to the end of accretion. This gives a natural explanation of 
the origin of those radio pulsars in binary systems which lie between 
millisecond pulsars and the main pulsar population in the B-P plane.

\section*{Acknowledgement}

This work was financially supported in part by the Russian Foundation
of Basic Research under the Grant 97-02-18096(a).

\section*{References}

Alpar M.A., Cheng A.F., Ruderman M., Shaham J. 1982, Nature 300, 728 \\
Bhattacharya D., van den Heuvel E.P.J. 1991, Physics Reports 203, 1 \\
Bhattacharya D., Srinivasan G. 1991, in "Neutron Stars: Theory and
Observations" (eds. J.Ventura \& D.Pines), Kluwer, Dordrecht, p.219 \\
Bondi H. 1952, MNRAS 112, 195 \\
Fujimoto M., Hanava T., Icko Iben Jr., Richardson M. 1984, ApJ 278, 813 \\
Geppert U., Urpin V. 1994, MNRAS 271, 490 \\
Geppert U., Urpin V., Konenkov D. 1995, A\&A 307, 807 \\
Goldreich P., Julian W. 1969, ApJ 157, 869 \\
Haensel P., Zdunik J.L. 1990, A\&A 227, 431 \\
Illarionov A., Sunyaev R. 1975, A\&A 39, 185 \\
Itoh N., Hayashi H., Kohyama Y. 1993, ApJ 418, 405 \\
Jahan Miri M., Bhattacharya D. 1994, MNRAS 289, 455 \\
Miralda-Escude J., Haensel P., Paczynski B. 1990, ApJ 362, 572 \\
Ostriker J., Gunn J. 1969, ApJ 157, 1395 \\
Pandharipande V., Smith R. 1975, Nucl.Phys. A237, 507 \\
Pringle J.E., Rees M.J. 1972, A\&A 21, 1 \\
Sang Y., Chanmugam G. 1987, ApJ 323, L61 \\
Taam R., van den Heuvel E.P.J. 1986, ApJ 305, 235 \\
Taylor J., Manchester R., Lyne A. 1993, ApJS 88, 529 \\
Urpin V., Geppert U. 1995, MNRAS 275, 1117 \\
Urpin V., Konenkov D. 1997, MNRAS, 292, 167, astro-ph/9801077 \\
van den Heuvel E.P.J., Bitzaraki O. 1995, A\&A 297, L41 \\
Van Riper K.A. 1991, ApJS 75, 449 \\
Yakovlev D., Urpin V. 1980, SvA 24, 303 \\

\newpage
\begin{figure}[t]
\vspace{20cm}
\input epsf
\epsfbox[ 70 140 0 0]{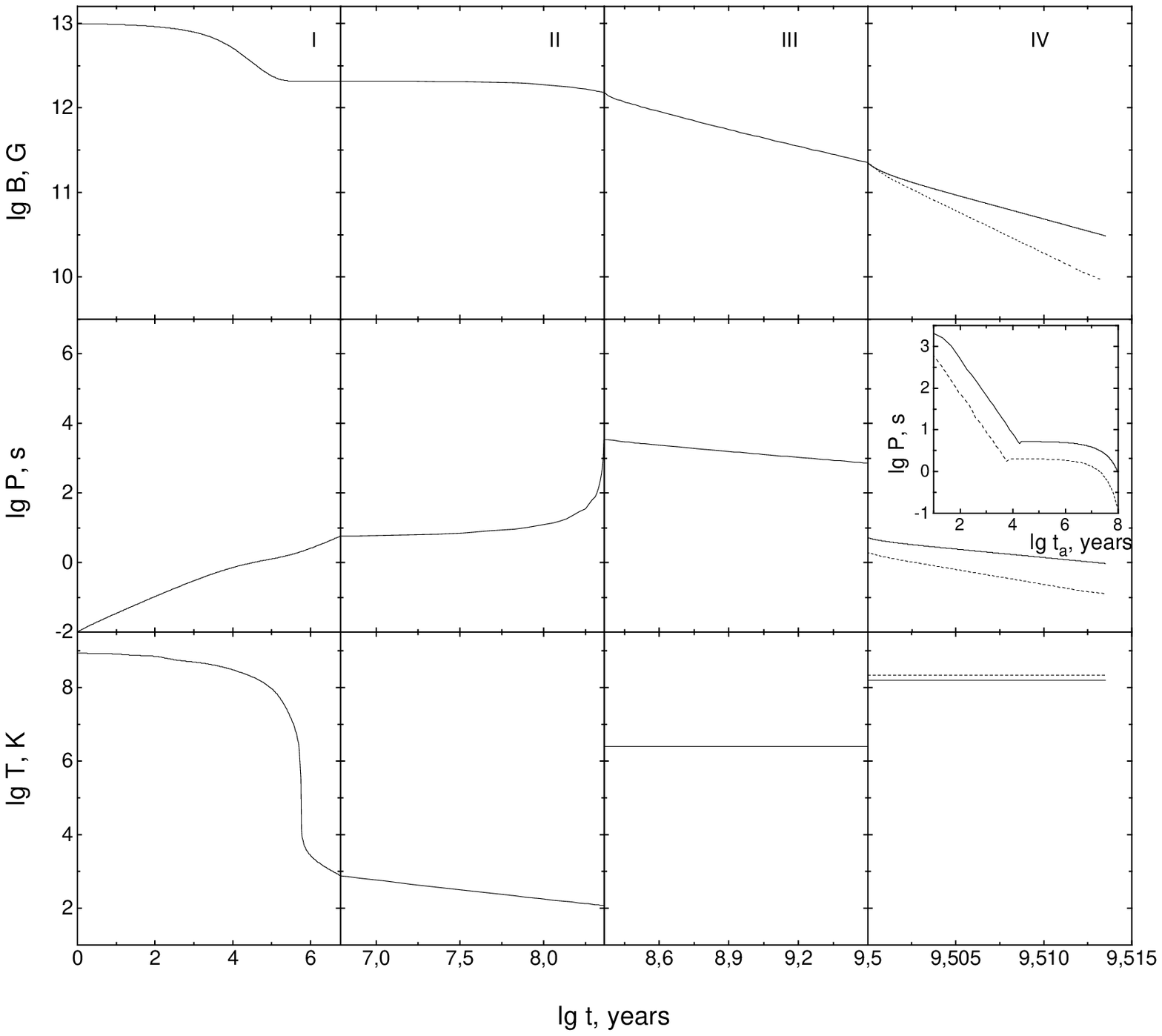}
\caption{The magnetic, spin and thermal evolution of the
neutron star with $B_{0}= 10^{13}$ G, $\rho_{0}= 10^{13}$ g/cm$^{3}$
and $Q=0.01$. The duration of the main-sequence lifetime of the companion
is $t_{ms} = 3 \times 10^{9}$ yr, the rate of accretion from the stellar 
wind is $10^{-15} M_{\odot}$/yr. The evolution is shown for two values of 
the accretion rate due to Roche-lobe overflow, $\dot{M} = 10^{-10}$ (solid 
lines) and $10^{-9} M_{\odot}$/yr (dotted lines). The duration of 
accretion is assumed to be $10^{8}$ yr for both cases. The panel shows
the spin evolution during the phase IV, $t_{a} = t - t_{ms}$.}
\end{figure}
\clearpage

\begin{figure}[t]
\vspace{20cm}
\input epsf
\epsfbox[ 10 140 0 0]{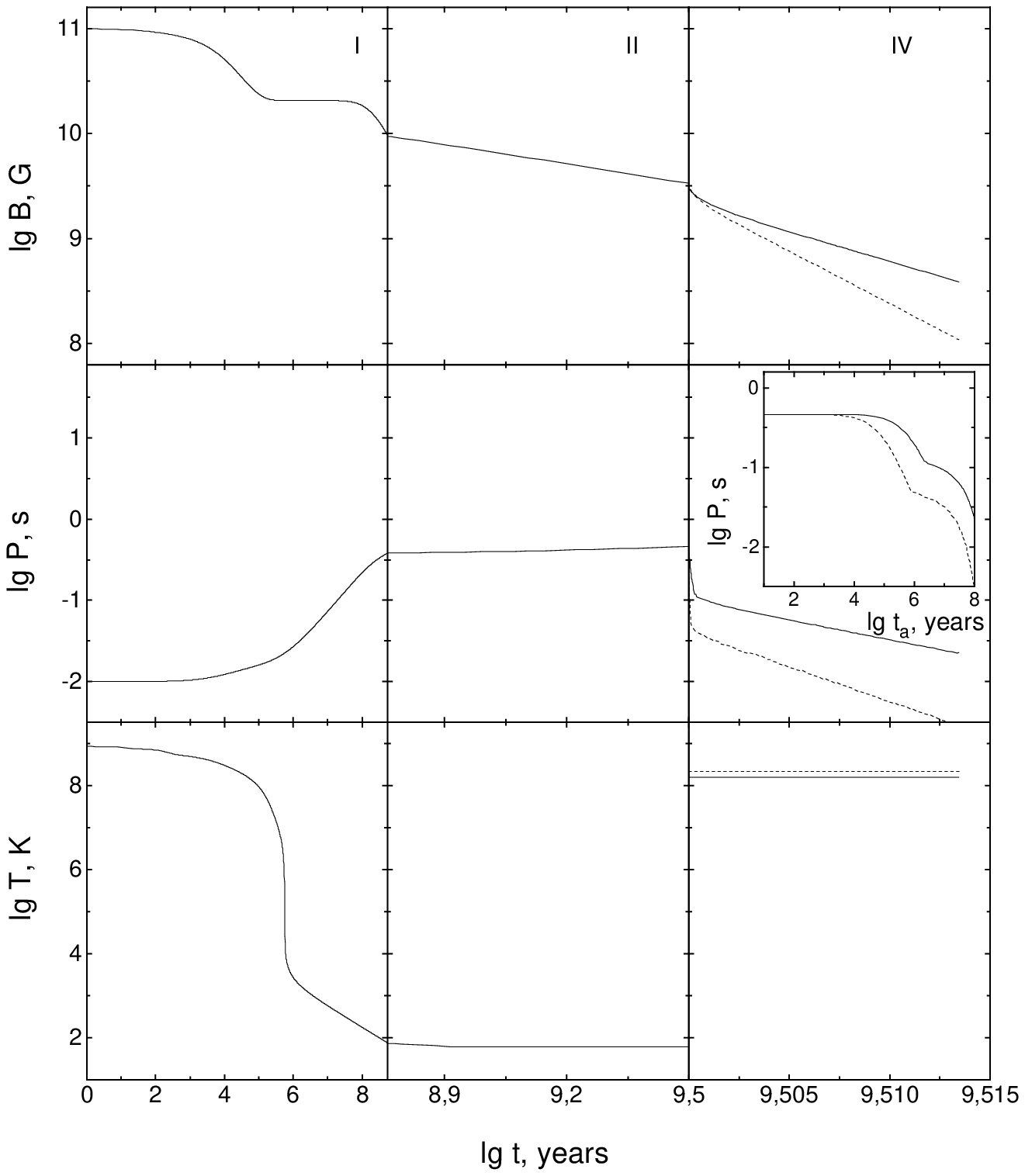}
\caption{Same as Fig.1 but for $B_{0}=10^{11}$ G.}
\end{figure}
\clearpage

\begin{figure}[t]
\vspace{20cm}
\input epsf
\epsfbox[ 70 140 0 0]{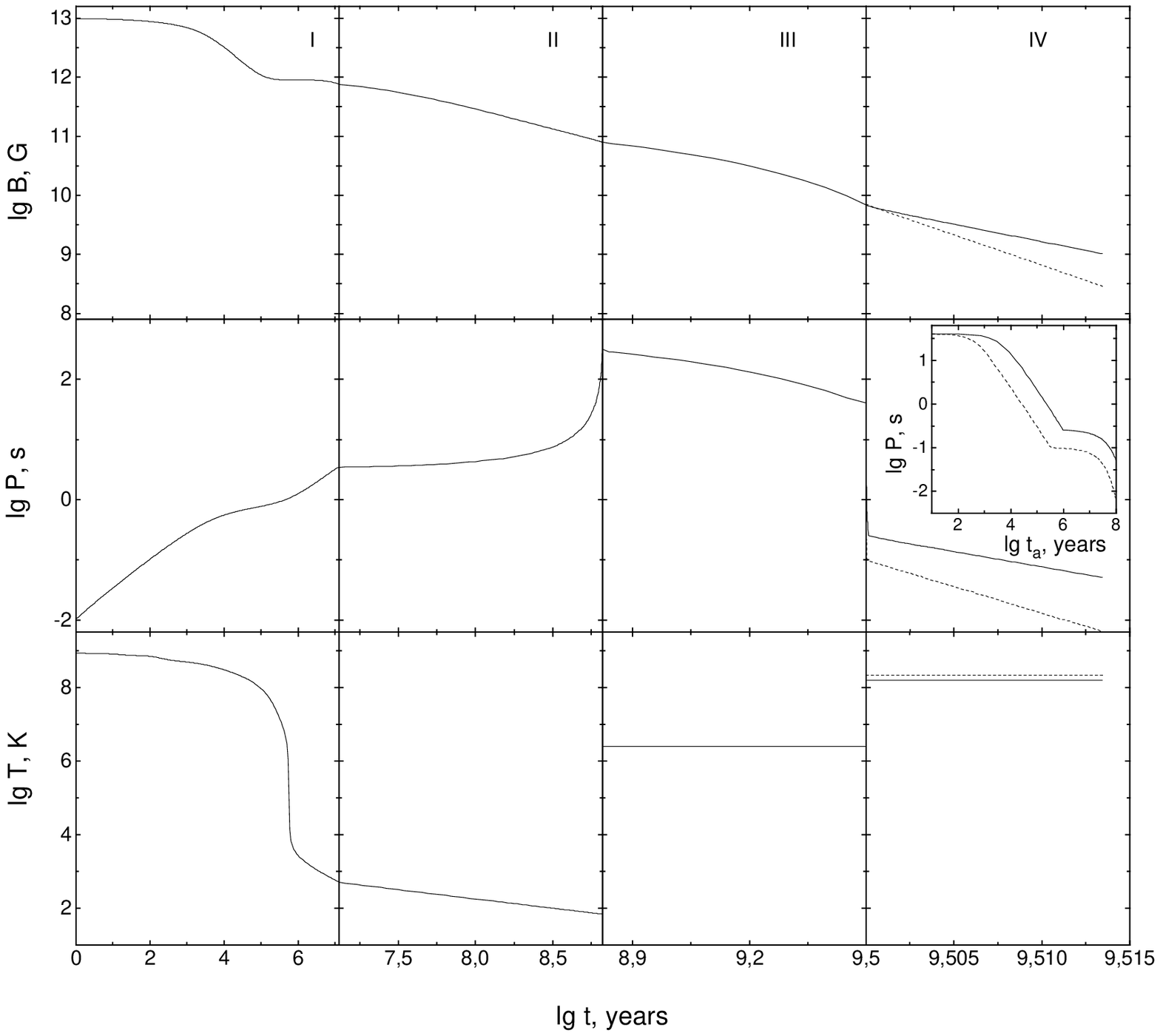}
\caption{Same as Fig.1. but for $\rho_{0} = 10^{12}$ g/cm$^{3}$
and $Q=0.1$.}
\end{figure}
\clearpage

\begin{figure}[t]
\vspace{20cm}
\input epsf
\epsfbox[ 70 140 0 0]{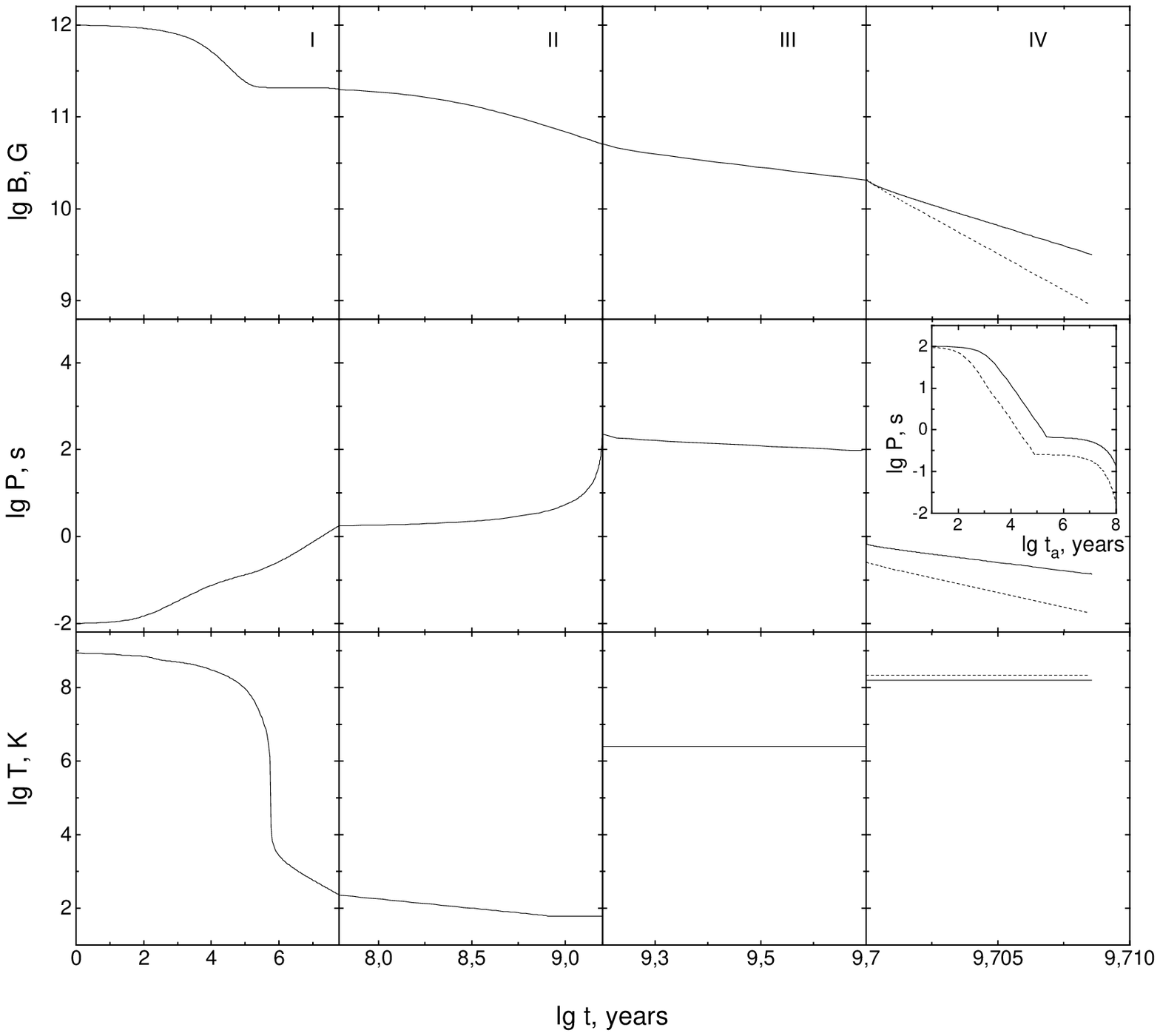}
\caption{Same as Fig.1. but for $B_{0}=10^{12}$ G and $t_{ms}=
5 \times 10^{9}$ yr.}
\end{figure}
\clearpage

\begin{figure}[t]
\vspace{20cm}
\input epsf
\epsfbox[ 70 100 0 0]{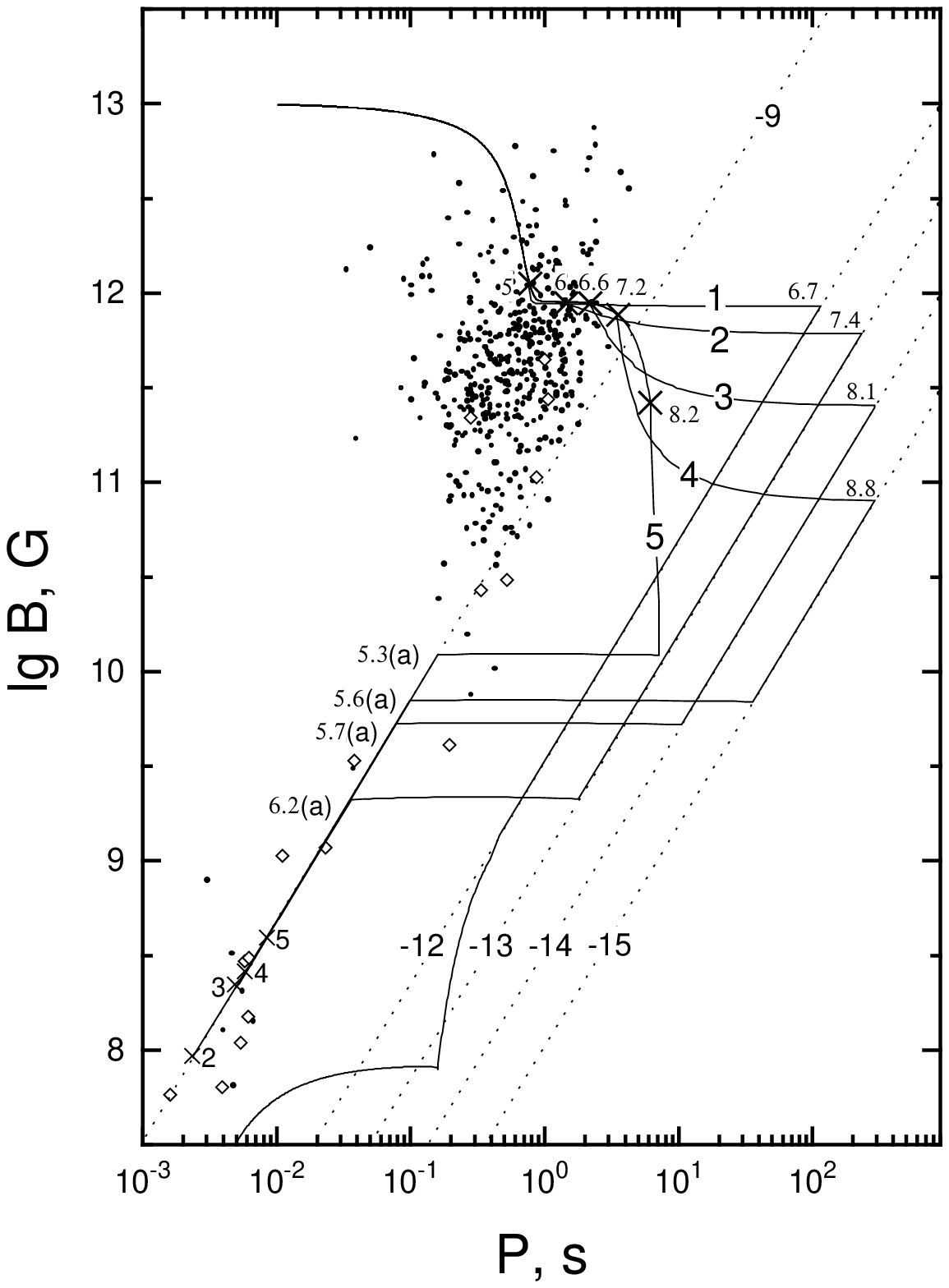}
\caption{The evolutionary tracks of the neutron star for
different rates of accretion from the stellar wind: $\dot{M} =
10^{-12}$ (curve 1), $10^{-13}$ (2), $10^{-14}$ (3), $10^{-15}$ (4) 
and $10^{-17} M_{\odot}$/yr (5). Other parameters are: $B_{0}=
10^{13}$ G, $\rho_{0}= 10^{12}$ g/cm$^{3}$, $Q=0.1$, $t_{ms} =
3 \times 10^{9}$ yr. The rate of accretion due to Roche-lobe
overflow is $10^{-9} M_{\odot}$/yr, the duration of this phase is
$10^{8}$ yr.}
\end{figure}
\clearpage

\begin{figure}[t]
\vspace{20cm}
\input epsf
\epsfbox[ 70 100 0 0]{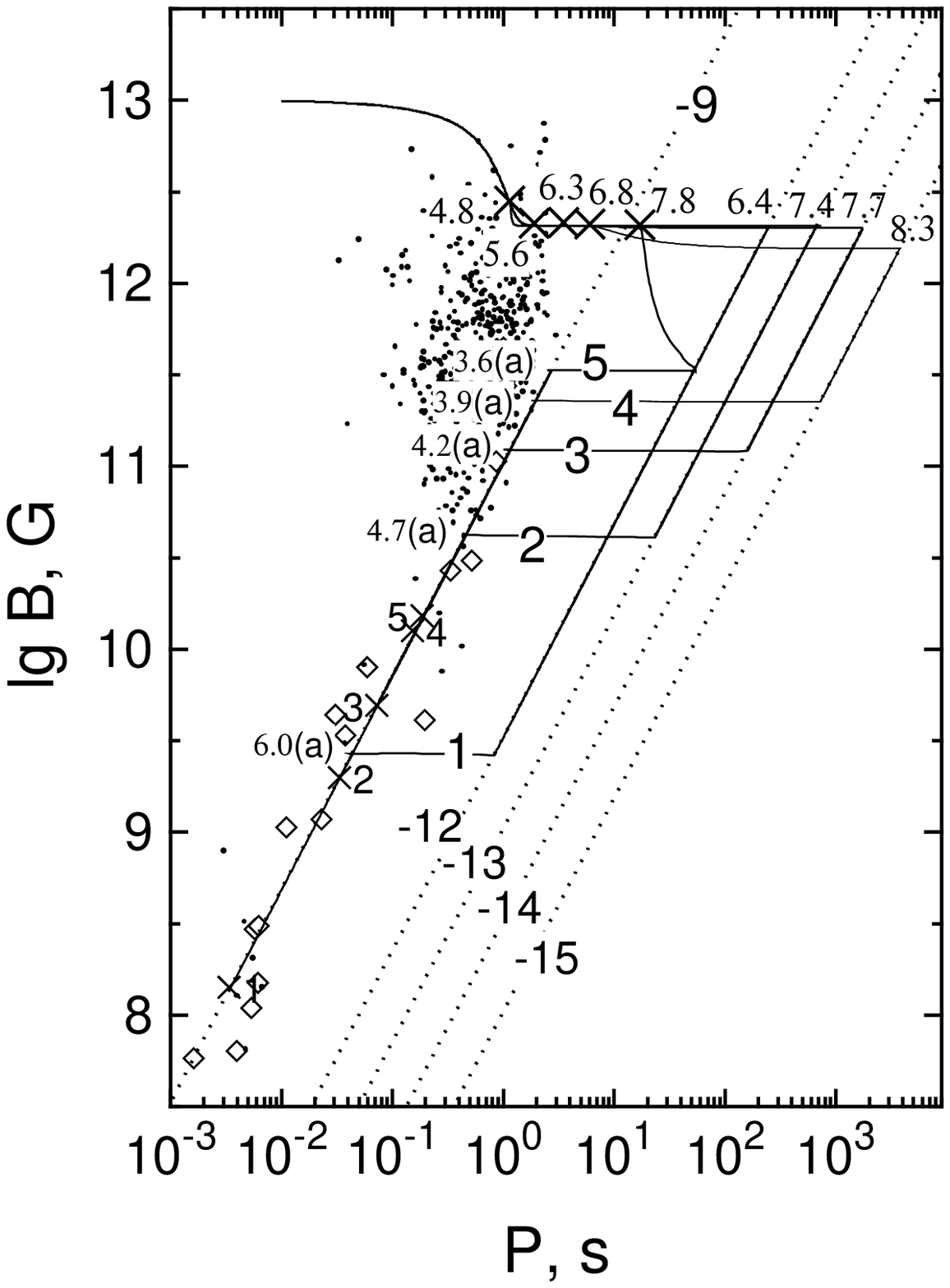}
\caption{Same as Fig.5 but for $\rho_{0}=10^{13}$ g/cm$^{3}$
and $Q=0.01$.}
\end{figure}
\clearpage

\begin{figure}[t]
\vspace{20cm}
\input epsf
\epsfbox[ 70 100 0 0]{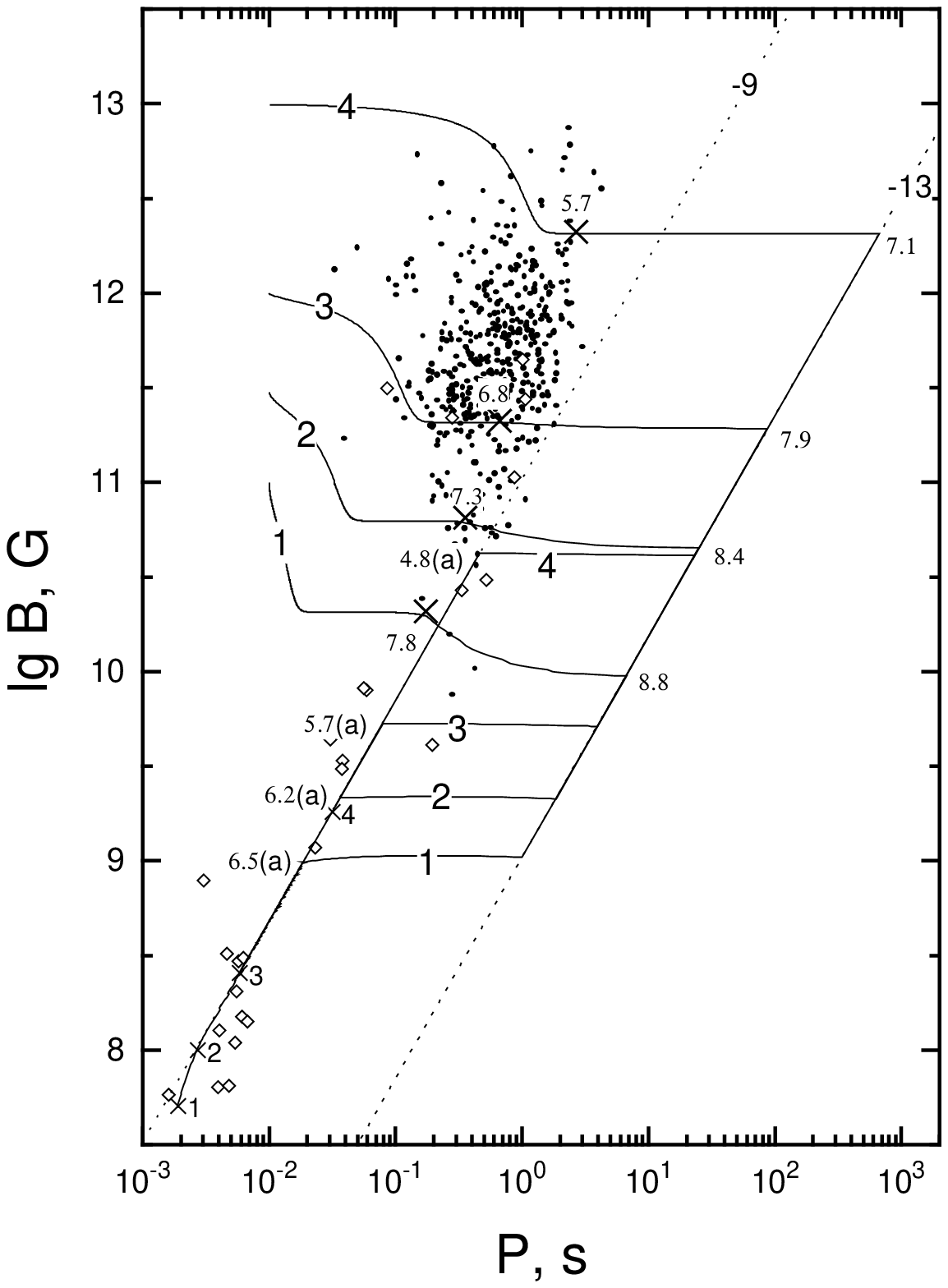}
\caption{The evolution of the neutron star with different 
initial magnetic fields: $B_{0}= 10^{11}$ (curve 1), $3 \times 10^{11}$
(2), $10^{12}$ (3) and $10^{13}$ G (4). Other parameters are: $\rho_{0}
=10^{13}$ g/cm$^{3}$, $Q=0.01$, $t_{ms} = 3 \times 10^{9}$ yr. The rate
of accretion from the stellar wind and due to Roche-lobe overflow is
$10^{-13}$ and $10^{-9} M_{\odot}$/yr, respectively.}
\end{figure}
\clearpage

\begin{figure}[t]
\vspace{20cm}
\input epsf
\epsfbox[70 100 0 0]{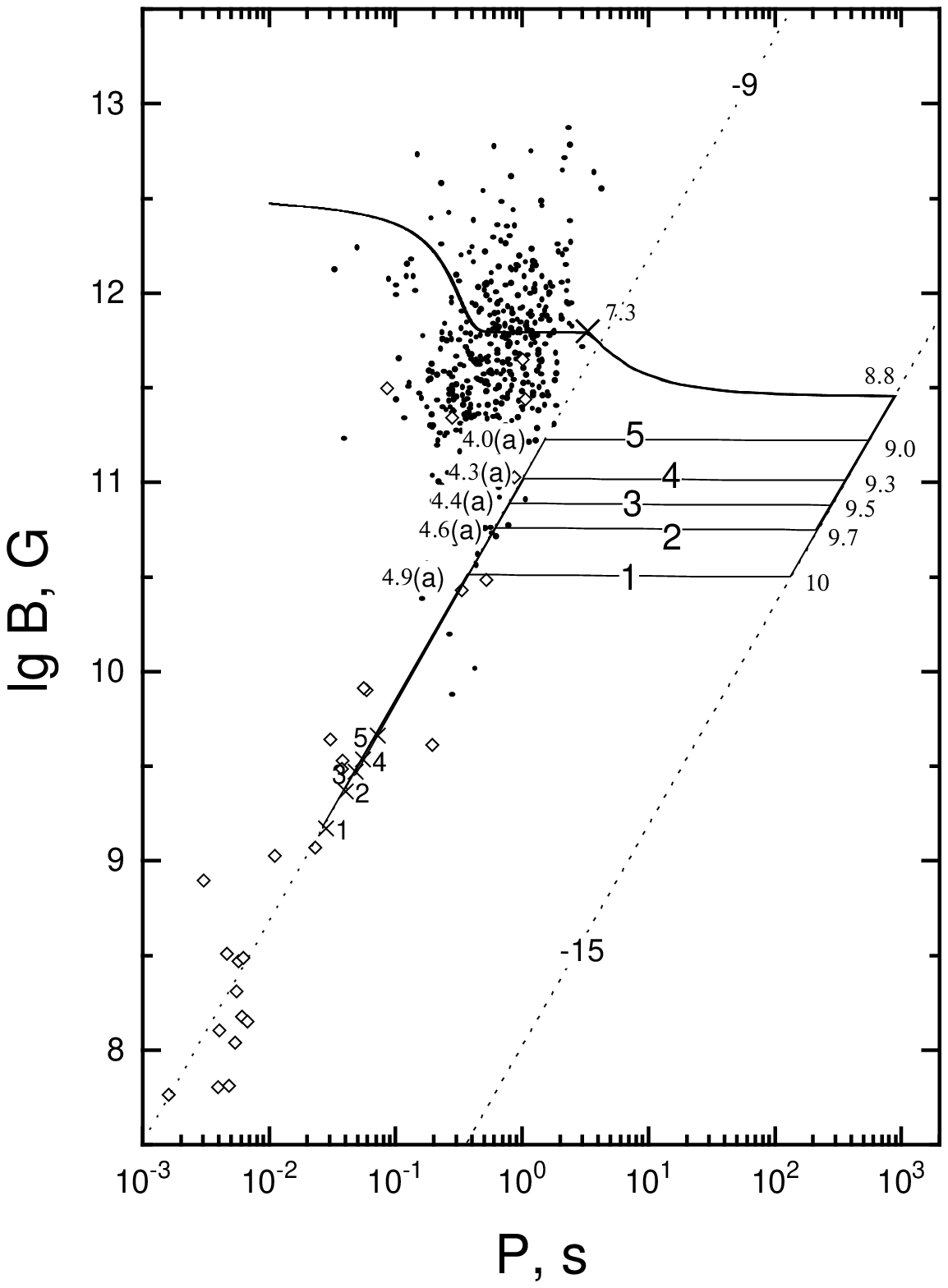}
\caption{The evolution of the neutron star in binary systems with
a different duration of the main-sequence lifetime of the companion: 
$t_{ms} = 10^{10}$ (curve 1), $5 \times 10^{9}$ (2), $3 \times 10^{9}$
(3), $2 \times 10^{9}$ (4) and $10^{9}$ yr (5). Other parameters are:
$\rho_{0}= 10^{13}$ g/cm$^{3}$, $B_{0}= 3 \times 10^{12}$ G, $Q=0.03$.
The rate of accretion from the stellar wind and due to Roche-lobe
overflow is $10^{-15}$ and $10^{-9} M_{\odot}$/yr, respectively.}
\end{figure}
\clearpage

\begin{figure}[t]
\vspace{20cm}
\input epsf
\epsfbox[70 100 0 0]{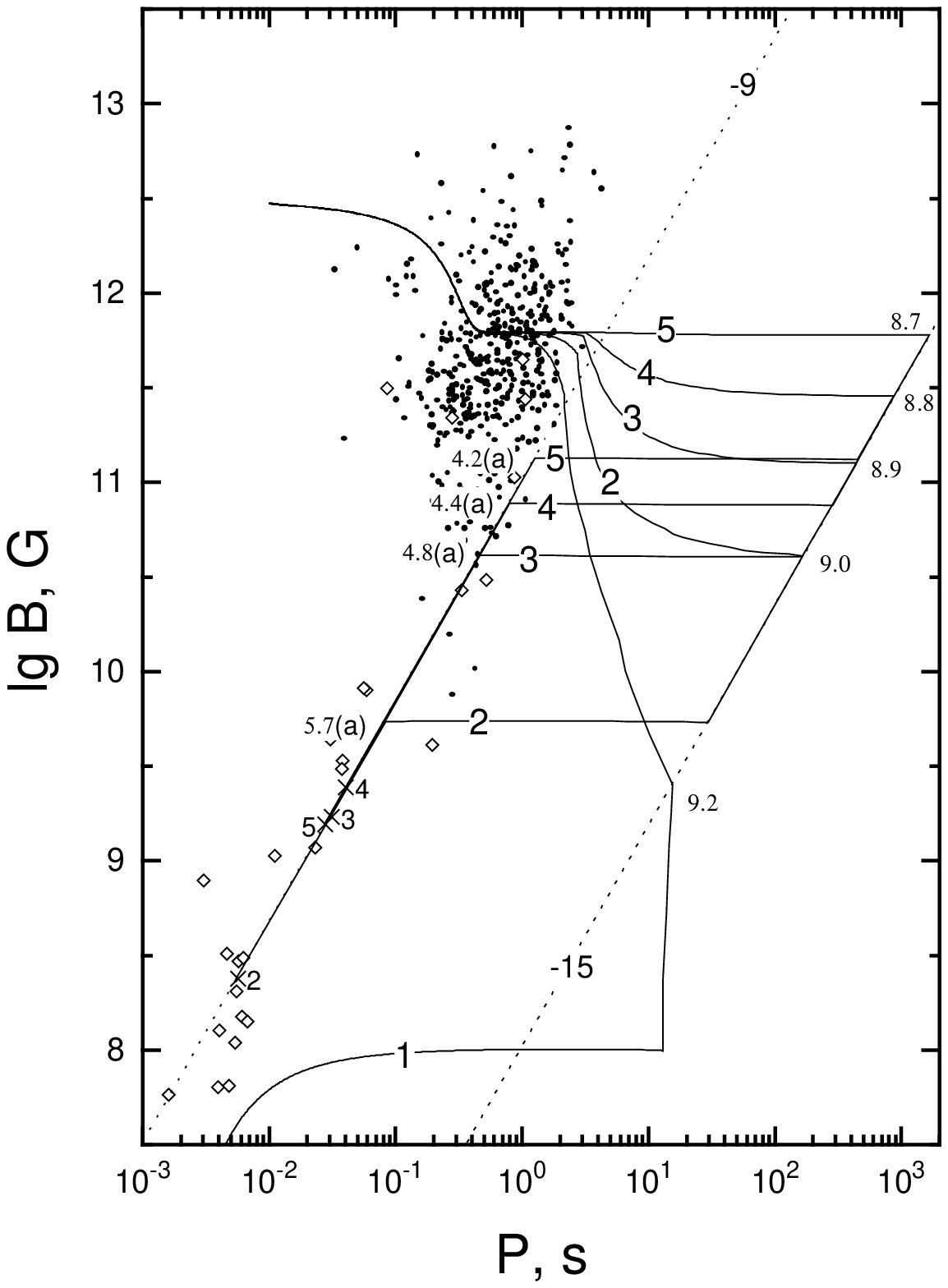}
\caption{The evolution of the neutron star with different values 
of the impurity parameter: $Q=0.3$ (curve 1), $0.1$ (2), $0.03$ (3),
$0.01$ (4) and $0.001$ (5). Other parameters are: $\rho_{0}=
10^{13}$ g/cm$^{3}$, $B_{0}= 3 \times 10^{12}$ G, $t_{ms} = 3 \times
10^{9}$ yr. The rate of accretion from the stellar wind and due to
Roche-lobe overflow is $10^{-15}$ and $10^{-9} M_{\odot}$/yr,
respectively.}
\end{figure}
\clearpage

\begin{figure}[t]
\vspace{20cm}
\input epsf
\epsfbox[70 100 0 0]{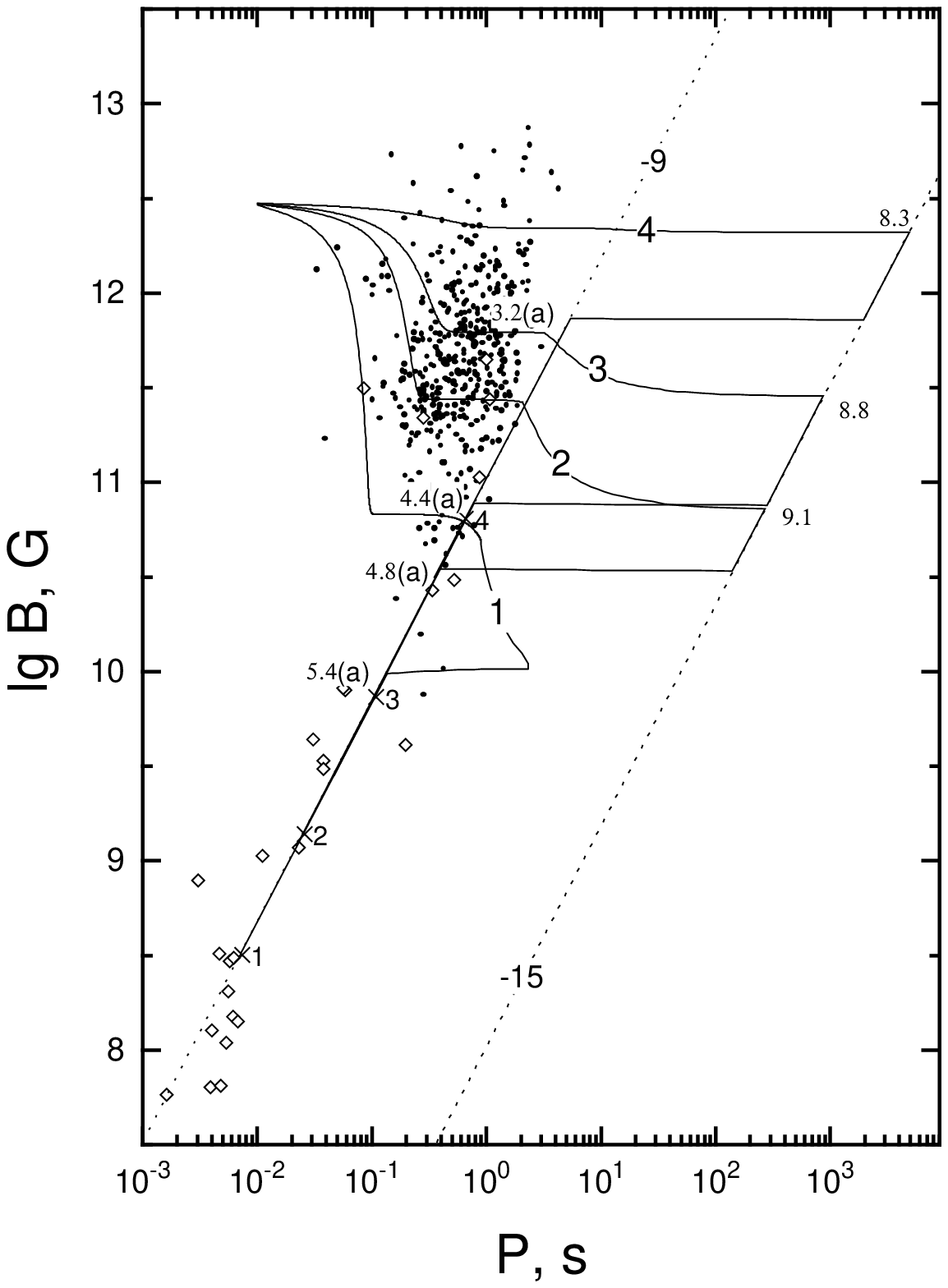}
\caption{The evolution of the neutron star with the original
magnetic field confined to the layers of a different density: $\rho_{0}=
10^{11}$ (curve 1), $10^{12}$ (2), $10^{13}$ (3) and $10^{14}$ 
g/cm$^{3}$ (4). Other parameters are: $B_{0}= 3 \times 10^{12}$ G, 
$t_{ms} = 3 \times 10^{9}$ yr, $Q=0.01$. The accretion rates are the 
same as in Fig.8 and 9.}
\end{figure}

\end{document}